\documentstyle[preprint,aps,prb,eqsecnum,tighten,epsbox]{revtex}

\def\beq{\begin{equation}}
\def\eeq{\end{equation}} 
\def\beqa{\begin{eqnarray}} 
\def\eeqa{\end{eqnarray}} 
\def\bfig{\begin{figure}\vspace{5mm}}

\def\efig{\end{figure}}

\def\bqu{\begin{quote}}
\def\equ{\end{quote}}
\def\bitem{\begin{itemize}}
\def\eitem{\end{itemize}}
\def\ben{\begin{enumerate}}
\def\een{\end{enumerate}}

\def\order{{\cal O}}

\def\<{\langle}
\def\>{\rangle}
\def\sbeqa{\begin{mathletters}}
\def\seeqa{\end{mathletters}}
\def\non{\nonumber}
\def\xis{\xi^{{\rm s}}}
\def\bxis{\bar{\xi}^{{\rm s}}}
\def\zs{z^{{\rm s}}}
\def\bzs{\bar{z}^{{\rm s}}}
\def\vap{\varphi}

\begin{document}
\draft
\title{
{\bf Macroscopic Quantum Dynamics of a Free Domain Wall 
in a Ferromagnet}}
\author{
Junya Shibata$^{\dag}$
 and Shin Takagi$^{\dag\dag}$}
\address{Department of Physics, Tohoku University, 
Sendai 980-8578, Japan}
\date{\today}
\maketitle
\bigskip
\begin{abstract}
We study macroscopic quantum dynamics of a free domain wall 
in a quasi-one-dimensional ferromagnet by use of 
the spin-coherent-state path integral in {\it discrete-time} formalism. 
Transition amplitudes between typical states are 
quantitatively discussed by use of {\it stationary-action approximation} 
with respect to collective degrees of freedom representing 
the center position and the chirality of the domain wall. 
It is shown that the chirality may be loosely said to be canonically 
conjugate to the center position; 
the latter moves with a speed depending on the former. 
It is clarified under what condition the center position 
can be regarded as an effective free-particle position, 
which exhibits the phenomenon of wave-packet spreading. 
We demonstrate, however, that in some case the non-linear character 
of the spin leads to such a dramatic phenomenon of a non-spreading 
wave packet as to completely invalidate the free-particle analogy. 
In the course of the discussion, 
we also point out various difficulties 
associated with the continuous-time formalism. 
\end{abstract}
\vspace{0.5cm}
\widetext

\section{Introduction}

Recent developments in low-temperature measurement techniques 
and the so-called nanostructure technology 
enable us to study low-dimensional magnetism in mesoscopic magnetic systems. 
Among others, dynamics of a domain wall in a ferromagnet has 
attracted much attention both theoretically and experimentally, 
because it is expected to exhibit quantum-mechanical aspects 
at sufficiently low temperatures.\cite{{S-C-B1},{QTM94},{C-T}}
A domain wall contains a (semi-)macroscopic number of spins, 
its width being typically $10$ $\sim$ $1000$ \AA. 
Hence, if its quantum-mechanical behavior was found, 
it would be an evidence of macroscopic quantum phenomena (MQP). 
To list just a few of the theoretical works about possible MQP 
involving such a domain wall: depinning of a domain wall via 
macroscopic quantum tunneling  (MQT),\cite{{Egami},{Stamp},{Chudnovsky},{Tatara-Fukuyama}} 
coherent tunneling though a periodic pinning potential,\cite{Braun-Loss} 
macroscopic quantum coherence (MQC) of the chirality.
\cite{{Braun-Loss},{Braun-Loss2},{Takagi-Tatara},{Ivanov}} 

One of the standard procedures to discuss MQP begins by deriving an 
effective action in terms of those collective degrees of freedom which 
directly describe the tunneling in question. 
In the case of the magnetic domain wall, the relevant collective degrees 
of freedom are the center position and/or the chirality of the wall. 
Existing works in the literature then treat the effective action 
in the Caldeira-Leggett scheme\cite{Caldeira-Leggett} to evaluate the tunneling rate. 
However, as emphasized by Leggett,\cite{Leggett} one should probe a quantum-mechanical time evolution to check whether MQP (especially, MQC) have occurred. 
Hence, what is needed on the theoretical side is to 
evaluate not only tunneling rates but also relevant transition amplitudes. 

As a technique to evaluate a transition amplitude, 
the spin-coherent-state path integral in {\it continuous-time} formalism
 \cite{Klauder} is frequently used. 
However, as noted by some workers,\cite{{Solari},{Kochetov}} 
this formalism has some fundamental difficulties 
even at the level of a single-spin system. 
The nature and implication of the difficulties 
have recently been examined in detail,\cite{Shibata-Takagi} 
where it has been pointed out, among others, that the information 
on the initial and the final states fails to be retained 
in the transition amplitude in question, and that, when combined with the 
{\it stationary-action approximation}, the stationary value of the action 
is not always given correctly and the fluctuation integral diverges.
Hence, at present, there are no reliable results for transition amplitudes. 
One of the interesting predictions made by the continuous-time formalism is 
an interference effect (the so-called spin-parity effect); 
the behavior of the domain wall is predicted to depend dramatically on 
whether the magnitude {\it S} of each of the individual spins 
is an integer or a half-integer. 
The origin of the effect has been ascribed to the so-called 
Berry-phase term appearing in the effective action.\cite{Braun-Loss} 
However, the information of the initial and the final states are 
essential for interference effects. 
Hence, the purported interference effect need be re-considered.

 In this paper, we present a first step to clarify these problems by use of 
the {\it discrete-time} formalism of the spin-coherent-state path integral. 
We focus upon a free domain wall and evaluate real-time transition amplitudes. 
This is the first case of an unambiguous evaluation of such amplitudes via 
the spin-coherent-state path integral as applied to 
an interacting many-spin system.

The paper is organized as follows. 
In Sec. II we present the model Hamiltonian to be treated, namely that 
consisting of the Heisenberg exchange and the anisotropy energies, 
and formulate transition amplitudes 
between spin coherent states 
by use of the spin-coherent-state path integral 
in the discrete-time formalism. 
In Sec. III we introduce a domain wall together with the 
collective degrees of freedom 
for its center position and 
the chirality, respectively. 
This section is also devoted to the 
derivation of the effective action of the domain wall.
We also point out some problems associated with 
the continuous-time treatment of 
the effective action. 
Sec. IV evaluates transition amplitudes for a 
free domain wall in the stationary-action approximation 
including the effects of fluctuations. 
In the course of the evaluation, 
we note the conjugate relation between the center position 
and the chirality. 
We also locate those terms which can induce interference effects. 
In Sec. V we compute transition probabilities between typical states 
and compare the quantum dynamics of the domain wall with that 
of a free particle. 
This comparison allows us, among others, to identify 
the "effective mass" of the domain wall. 
In Sec. VI it is explicitly shown that 
the continuous-time formalism leads to a wrong 
transition probability for a free domain wall. 
We conclude with a speculation 
on the possibility of MQT and/or MQC involving a domain wall.  

\section{Model}

We consider a ferromagnet consisting of a spin $\mbox{\boldmath$S$}$ of magnitude ${\it S}$ at each site in a quasi-one-dimensional cubic crystal 
(a linear chain) 
of lattice constant $a$. The magnet is assumed to have an easy axis and a hard axis in the $z$ and the $x$ directions, respectively. Accordingly, we adopt the Hamiltonian 
\beqa
\hat{H} = -\tilde{J}\sum_{<i,j>}^{N_{L}}\hat{\mbox{\boldmath$S$}}_{i}
\cdot\hat{\mbox{\boldmath$S$}}_{j} 
          -\frac{1}{2}\sum_{j}^{N_{L}}(K \hat{S}^2_{j,z}- K_{\bot}\hat{S}^2_{j,x}),
\label{hami1}
\eeqa
where the index $i$ or $j$ represents a lattice point, $<i,j>$ denotes 
a nearest-neighbor pair, $N_{L}$ is the total number of lattice points, 
and $\tilde{J}$ is the exchange coupling constant, 
and $K$ and $K_{\bot}$ are longitudinal and transverse anisotropy constants; 
$\tilde{J}$, $K$, and $K_{\bot}$ are all positive.

Since we are interested in those transition amplitudes 
which are appropriate to describe quantum mechanical motion of a domain wall, 
we introduce a {\it spin-coherent state}\cite{Radcliffe} at each site, 
which is suited for a vector picture of spin. 
By use of the eigenstate $|S\>$ of $\hat{S}_{z}$ associated 
with the eigenvalue ${\it S}$, the spin coherent state is defined by 
\beqa
|\mbox{\boldmath$n$}\> := (1+|\xi|^{2})^{-S}\exp(\xi\hat{S}_{-})|S\>,
\label{scs1}
\eeqa
where $\mbox{\boldmath$n$}$ is a unit vector $(n_{x}=\sin\theta\cos\phi,n_{y}=\sin\theta\sin\phi,n_{z}=\cos\theta)$ with the complex number $\xi$ 
being its Riemann projection:
\beqa
\xi= e^{i\phi}\tan\frac{\theta}{2},\qquad \xi^{*}= e^{-i\phi}\tan \frac{\theta}{2}.
\label{xi}
\eeqa
These states form an overcomplete set and possess, among others, the following properties:
\sbeqa
\label{scspro}
\beqa
\<\mbox{\boldmath$n$}|\hat{\mbox{\boldmath$S$}}|\mbox{\boldmath$n$}\> &=& S\mbox{\boldmath$n$},
\label{scspro1}\\
\<\<\mbox{\boldmath$n$}'|(\mbox{\boldmath$e$}\cdot\hat{\mbox{\boldmath$S$}})^{2}|\mbox{\boldmath$n$}\>\> &:=& \frac{\<\mbox{\boldmath$n$}'|(\mbox{\boldmath$e$}\cdot\hat{\mbox{\boldmath$S$}})^{2}|\mbox{\boldmath$n$}\>}{\<\mbox{\boldmath$n$}'|\mbox{\boldmath$n$}\>}\non\\
&=& \left(1-\frac{1}{2S}\right)
(\<\<\mbox{\boldmath$n$}'|\mbox{\boldmath$e$}\cdot\hat{\mbox{\boldmath$S$}}
|\mbox{\boldmath$n$}\>\>)^{2} + \frac{S}{2},
\label{scspro2}
\eeqa
\seeqa 
where $\mbox{\boldmath$e$}$ is an arbitrary unit vector. 
Hereafter we work with the $\xi$-representation, 
and denote a state of the system as 
\beqa
|\xi\> \equiv |\xi_{1},\xi_{2},\cdots,\xi_{N_{L}}\>
        := \bigotimes_{j}^{N_{L}}|\xi_{j}\>,
\label{total-xi}
\eeqa
where $|\xi_{j}\>(\equiv |\mbox{\boldmath$n$}_{j}\>$) is 
a spin coherent state at the site $j$.
The transition amplitude between the initial state $|\xi_{{\rm I}}\>$ and the final 
state $|\xi_{{\rm F}}\>$ can be expressed as a 
{\it spin-coherent-state path integral} in the real {\it discrete-time} 
formalism by the standard procedure of the repeated use of the resolution 
of unity (see, e.g., Ref. 17 on which the present notation is based):
\beqa
\< \xi_{{\rm F}}|e^{-i\hat{H}T/\hbar}|\xi_{{\rm I}} \>
= \lim_{N\to\infty}\int\prod_{n=1}^{N-1}\prod_{j}^{N_{L}}
d\mu(\xi_{j}(n),\xi^{*}_{j}(n))
\exp\left(\frac{i}{\hbar}{\cal S}[\xi^{*},\xi]\right),
\label{ta1}
\eeqa
where $N \equiv T/\epsilon$, $\epsilon$ is an infinitesimal time interval, 
$n$ represents discrete time, and the integration measure is 
\beqa
d\mu(\xi_{j}(n),\xi^{*}_{j}(n)) 
:= \frac{2S+1}{(1+|\xi_{j}(n)|^2)^2}\frac{d\xi_{j}(n)d\xi^{*}_{j}(n)}{2\pi i },
\qquad \frac{d\xi_{j}(n)d\xi^{*}_{j}(n)}{2\pi i }\equiv \frac{d\Re{\xi}_{j}(n)d\Im{\xi}_{j}(n)}{\pi}.
\label{measure1}
\eeqa
The action ${\cal S}[\xi^{*},\xi]$ consists of two parts, 
${\cal S}^{{\rm c}}[\xi^{*},\xi]$ and ${\cal S}^{{\rm d}}[\xi^{*},\xi]$, 
which are to be called the {\it canonical term} and the {\it dynamical term}, respectively. They take the following forms:
\sbeqa
\label{scs}
\beqa
&&{\cal S}[\xi^{*},\xi] = {\cal S}^{{\rm c}}[\xi^{*},\xi] + {\cal S}^{{\rm d}}[\xi^{*},\xi],
\label{scs-c-d}\\
&&\frac{i}{\hbar}{\cal S}^{{\rm c}}[\xi^{*},\xi] 
:= \sum_{n=1}^{N}\sum_{j}^{N_{L}}\ln\<\xi_{j}(n)|\xi_{j}(n-1)\>\non\\
&&=S \sum_{n=1}^{N}\sum_{j}^{N_{L}}
\ln \frac{(1+\xi^{*}_{j}(n)\xi_{j}(n-1))^2}{(1+|\xi_{j}(n)|^2)(1+|\xi_{j}(n-1)|^2)},
\label{scs-c}\\
&&\frac{i}{\hbar}{\cal S}^{{\rm d}}[\xi^{*},\xi]
:= -\frac{i}{\hbar}\sum_{n=1}^{N}\epsilon H(\xi^{*}(n),\xi(n-1)),
\label{scs-d}\\
&&H(\xi^{*},\eta) := \<\< \xi|\hat{H}| \eta \>\>,
\qquad \xi(0)\equiv\xi_{{\rm I}},\qquad\xi(N)\equiv\xi_{{\rm F}}.
\eeqa
\seeqa
Here, we emphasize that the integration variables are 
$ \{\xi^{*}(n),\xi(n)| n=1,2,...,N-1\}$; 
$\xi(0)$ and $\xi(N)$ are fixed complex numbers. 
In passing, note that 
\beqa
\<\xi_{j}(n)|\xi_{j}(n-1)\>=\left(
\cos^{2}\frac{\theta_{j}(n)}{2}\cos^{2}\frac{\theta_{j}(n-1)}{2}
+\sin^{2}\frac{\theta_{j}(n)}{2}\sin^{2}\frac{\theta_{j}(n-1)}{2}
e^{-i(\phi_{j}(n)-\phi_{j}(n-1))}
\right)^{2S},
\label{peri-c}
\eeqa
which is a $2\pi$-periodic function of the phase difference 
$\phi_{j}(n)-\phi_{j}(n-1)$.

 At this stage, if one regarded all the differences $|\xi_{j}(n)-\xi_{j}(n-1)|$ as small in some sense, expanded the action ${\cal S}[\xi^{*},\xi]$ and went over to the {\it continuous-time} formalism, one would obtain 
the following form which had been used in most of the literature 
including Refs. 8 and 10:
\sbeqa
\label{conti-ta}
\beqa
&&\< \xi_{{\rm F}}|e^{-i\hat{H}T/\hbar}|\xi_{{\rm I}} \>
 \sim \int {\cal D}\xi{\cal D}\xi^{*}\exp
\left\{\frac{i}{\hbar}\left({\cal S}^{{\rm c}}_{{\rm con}}[\xi^{*},\xi] + 
{\cal S}^{{\rm d}}_{{\rm con}}[\xi^{*},\xi]\right)\right\},
\label{conti-ta1}\\
&&\frac{i}{\hbar}{\cal S}^{{\rm c}}_{{\rm con}}[\xi^{*},\xi]
:= S\sum_{j}^{N_{L}}\int_{0}^{T}dt
\frac{\dot{\xi}_{j}^{*}(t)\xi_{j}(t)-\xi^{*}_{j}(t)\dot{\xi}_{j}(t)}{1+\xi_{j}^{*}(t)\xi_{j}(t)},
\label{cnoti-canonical1}\\
&&\frac{i}{\hbar}{\cal S}^{{\rm d}}_{{\rm con}}[\xi^{*},\xi]
:=-\frac{i}{\hbar}\int_{0}^{T}dt H(\xi^{*}(t),\xi(t)),
\label{conti-dynamical1}
\eeqa
\seeqa 
where ${\cal D}\xi{\cal D}\xi^{*}$ represents 
a symbolic measure in the continuous-time formalism, 
However, as pointed out in our previous paper,\cite{Shibata-Takagi} 
this formalism has various difficulties (see also the next section).
For this reason, we proceed to consider the transition amplitude 
in the discrete-time formalism (\ref{scs}).

We shall be interested in 
those spin configurations whose scale of spatial variation 
is much larger than the lattice constant $a$. 
Accordingly, 
we take the spatial continuum limit in (\ref{scs-c}) and (\ref{scs-d}):
\sbeqa
\label{sconti-action}
\beqa
\frac{i}{\hbar}{\cal S}^{{\rm c}}[\xi^{*},\xi]
&=& S\sum_{n=1}^{N}\int_{-L/2}^{L/2}\frac{dx}{a}
\ln\frac{(1+\xi^{*}(x,n)\xi(x,n-1))^2}{(1+|\xi(x,n)|^2)(1+|\xi(x,n-1)|^2)},
\label{sconti-canonical}\\
\frac{i}{\hbar}{\cal S}^{{\rm d}}[\xi^{*},\xi]
&=&-\frac{i}{\hbar}\sum_{n=1}^{N}\epsilon \int_{-L/2}^{L/2}\frac{dx}{a}
{\cal H}(\xi^{*}(x,n),\xi(x,n-1)),
\label{sconti-dynamical}\\
{\cal H}(\xi^{*}(x),\eta(x)) 
&:=&\frac{S}{(1+\xi^{*}(x)\eta(x))^{2}}
\Bigg[ 2JS \partial_{x} \xi^{*}(x)\partial_{x}\eta(x)\non\\
&&-\frac{K}{2}\left\{\left(S-\frac{1}{2}\right)(1-\xi^{*}(x)\eta(x))^{2}
+\frac{1}{2}\right\}\non\\
&&+\frac{\alpha K}{2}\left\{\left(S-\frac{1}{2}\right)(\xi^{*}(x)+\eta(x))^{2}+\frac{1}{2}\right\}\Bigg],\label{sconti-hami}
\eeqa
\seeqa 
where $L$ is the length of the linear chain, 
$J \equiv \tilde{J}a^{2}$, 
and $\alpha \equiv K_{\bot}/K$. 
In this paper we consider the case of a weak transverse anisotropy 
$\alpha \ll 1$, 
and study the dynamics of a domain wall to the first order in $\alpha$.

\section{Effective Action for a Domain Wall}
\subsection{Kink configuration}

We begin by finding a domain wall configuration. 
It is determined by one of the static solutions $\{ \xis(x),\bxis(x)\}$ 
of the action ${\cal S}[\xi^{*},\xi]$.
They satisfy the following equations up to $\order(\alpha^{0})$:
\sbeqa
\label{stat-eq}
\beqa
&&\lambda^{2}\left\{\partial_{x}^{2}\xis(x)-\frac{2\bxis(x)(\partial_{x}\xis(x))^2}{1+\bxis(x)\xis(x)}\right\}-\frac{1-\bxis(x)\xis(x)}{1+\bxis(x)\xis(x)}\xis(x) =0,
\label{stat-eq1}\\
&&\lambda^{2}\left\{\partial_{x}^{2}\bxis(x)-\frac{2\xis(x)(\partial_{x}\bxis(x))^2}{1+\bxis(x)\xis(x)}\right\}-\frac{1-\bxis(x)\xis(x)}{1+\bxis(x)\xis(x)}\bxis(x) =0,
\label{stat-eq2}
\eeqa
\seeqa 
where $\lambda^2 \equiv JS/K(S-1/2)$. 
An obvious solution is the "vacuum" solution representing 
the uniform configuration in which the spins are 
either all parallel or all anti-parallel to the $z$ direction. 
The other solution is the "kink" solution representing
a domain-wall configuration in which the spins at $x\sim +\infty$ 
are parallel to the $z$ direction, the spins at $x\sim -\infty$ 
are anti-parallel to the $z$ direction, and there is a transition region 
(i.e., a domain wall) of width $\lambda$; 
\beqa
\xis(x) = \exp\left(-\frac{x-Q}{\lambda} + i\phi_{0}\right),
\qquad
\bxis(x) = \exp\left(-\frac{x-Q}{\lambda} - i\phi_{0}\right),
\label{stat-sol}
\eeqa
where $Q$ and $\phi_{0}$ are arbitrary real constants. 
$Q$ is the center position of the domain wall, 
and $\phi_{0}$ is a quantitative measure of the {\it chirality} 
of the domain wall with respect to the $x$ axis (Fig. 1); 
the wall is maximally right-handed if $\phi_{0} = \pi/2$ 
and maximally left-handed if $\phi_{0}=-\pi/2$, 
while it has no chirality if $\phi_{0}=0$. 
The range of $\phi_{0}$ is chosen as $-\pi \le \phi_{0} \le \pi$, 
with $\phi_{0}=\pi$ and $\phi_{0}=-\pi$ representing the same situation. 
$\{\xis(-x), \bxis(-x)\}$ is also a solution representing 
a domain-wall configuration. 
However, this as well as the vacuum solution 
belongs to a sector different from that of (\ref{stat-sol}). 
Since a transition 
between different sectors are forbidden,\cite{Rajaraman} 
it is sufficient to consider only the sector 
(\ref{stat-sol}) for the purpose of studying
 the dynamics of a domain wall. 

\subsection{Collective degrees of freedom}

Study of the domain-wall dynamics 
is facilitated by introducing relevant collective degrees of freedom. 
We note two kinds of invariance possessed by (\ref{stat-eq}). 
One is the translation invariance in the $x$ direction.  
The other is the rotation invariance around the $z$ axis.  
These invariances are embodied by the arbitrariness in the choice of $Q$ 
and $\phi_{0}$, respectively, in (\ref{stat-sol}). 
Hence we elevate them to dynamical variables $Q(n)$  
and $\phi_{0}(n).$\cite{{Braun-Loss},{Takagi-Tatara},{Rajaraman}}
To deal with these two dynamical variables (collective degrees of freedom), 
it is convenient to define 
\beqa
&&z(n) :=  q(n) + i\phi_{0}(n), \qquad 
z^{*}(n) := q(n) - i\phi_{0}(n),
\label{z-qp}\\
&&q(n)\equiv Q(n)/\lambda,\qquad n = 1,2,...,N-1. \nonumber
\eeqa
By use of these variables, original integration variables 
$\xi(x,n)$ and $\xi^{*}(x,n)$ may be decomposed 
into the domain-wall configuration and the environment around it:
\sbeqa
\label{xi-cf}
\beqa
\xi(x,n) &=& \xis(x;z(n)) + \eta(x,n;z(n)),\\
\label{xi-cf1}
\xi^{*}(x,n) &=& \bxis(x;z^{*}(n)) + \eta^{*}(x,n;z^{*}(n)),
\label{xi-cf2}
\eeqa
\seeqa
where
\beqa
\xis(x;z(n)) := \exp\left(-x/\lambda + z(n)\right) ,\qquad
\bxis(x;z^{*}(n)) := \exp\left(-x/\lambda + z^{*}(n)\right).
\eeqa
Hereafter we consider transition amplitudes between the 
following domain-wall states:
\beqa
|\xi_{\beta}\>=|z_{\beta}\>
:=\bigotimes_{j}^{N_{L}}|\xis(ja;z_{\beta})\>,\qquad \beta=I,F, 
\eeqa
where $z_{{\rm I}}$ represents the center position $q_{{\rm I}}$ 
and the chirality $\phi_{{\rm I}}$ of the domain wall in the initial state, and 
$z_{{\rm F}}$ those in the final state. 
At both ends of the discrete time ($n=0$ or $n=N$), we define
\sbeqa
\label{def0N}
\beqa
&&z(0) \equiv z_{{\rm I}} := q_{{\rm I}}+i\phi_{{\rm I}},\qquad 
z(N) \equiv z_{{\rm F}} := q_{{\rm F}}+i\phi_{{\rm F}},
\label{z0N}\\
&&\eta(x,0;z(0))=\eta^{*}(x,N;z^{*}(N))=0.
\label{eta0N}
\eeqa
\seeqa

Putting Eqs. (\ref{xi-cf}) into the action (\ref{sconti-action}), 
we obtain up to $\order(\alpha)$
\sbeqa
\label{scs-s}
\beqa
{\cal S}[\xi^{*},\xi]&=&
{\cal S}^{{\rm s}}[z^{*},z] + {\rm terms~involving~the~environment}~\eta, 
\label{action1}\\
{\cal S}^{{\rm s}}[z^{*},z] &=& {\cal S}^{{\rm sc}}[z^{*},z]+{\cal S}^{{\rm sd}}[z^{*},z],
\label{scs-sc-sd}\\
\frac{i}{\hbar}{\cal S}^{{\rm sc}}[z^{*},z]
&:=& \frac{N_{{\rm DW}}S}{2}\sum_{n=1}^{N}\int_{-L/\lambda}^{L/\lambda}dx \ln 
\frac{(1+e^{-x+z^{*}(n)+z(n-1)})^2}{(1+e^{-x+z^{*}(n)+z(n)})(1+e^{-x+z^{*}(n-1)+z(n-1)})},
\label{scs-sc}\\
\frac{i}{\hbar}{\cal S}^{{\rm sd}}[z^{*},z]&:=&
-\frac{i}{\hbar}E_{{\rm DW}}\sum_{n=1}^{N}\epsilon 
\left[ 1 + \frac{\alpha}{4}\left\{1+\cosh(z^{*}(n)-z(n-1))\right\} \right],
\label{scs-sd}
\eeqa
\seeqa 
where $N_{{\rm DW}}\equiv \lambda /a$ is the number of spins 
in the domain wall, and $E_{{\rm DW}} \equiv 2N_{{\rm DW}}KS(S-1/2)$ is the kink energy. 
The zero point of energy has been adjusted in (\ref{scs-sd}). 
In this paper, we consider only ${\cal S}^{{\rm s}}[z^{*},z]$ 
which is expected to make the most dominant contribution to 
transition amplitudes. 
The influence of the environment 
shall be discussed in a separate paper. 

Expression (\ref{scs-sc}) can be reduced to 
a simpler form (see Appendix A for the details of the derivation 
by use of {\it dilogarithm}\cite{Abramowitz-Stegun}):
\beqa
&&\frac{i}{\hbar}{\cal S}^{{\rm sc}}[z^{*},z]/N_{{\rm DW}}S\non\\
&&=\frac{1}{2}\sum_{n=1}^{N}\bigg[
-(q(n)-q(n-1))^{2}-R(\phi_{0}(n)-\phi_{0}(n-1))\non\\
&&-i\{2(q(n)+q(n-1))+2L/\lambda\}I(\phi_{0}(n)-\phi_{0}(n-1))
\bigg],
\label{red-scs-s}
\eeqa
where both $R(\phi)$ and $I(\phi)$ are $2\pi$-periodic functions (see, Fig. 2) 
such that 
\sbeqa
\label{periodic-func}
\beqa
R(\phi)&=& \phi^{2}~~~~:|\phi|\le\pi,\\
I(\phi)&=& \phi ~~~~~:-\pi \le \phi < \pi.
\label{RI}
\eeqa
\seeqa
This periodicity follows inevitably from the periodicity of (\ref{scs-sc}). 
In general this periodicity need be respected in performing 
the integration with respect to $\{\phi_{0}(n)\}$. 
However, in special circumstances when 
the range of $\phi_{0}(n)-\phi_{0}(n-1)$ can be restricted to 
$[-\pi,\pi]$ for all $n$, the above action may be rewritten as
\beqa
&&\frac{i}{\hbar}{\cal S}^{{\rm sc}}[z^{*},z]/N_{{\rm DW}}S\non\\
&=&\sum_{n=1}^{N}\bigg\{
-\frac{1}{2}\left((q(n)-q(n-1))^{2}+(\phi_{0}(n)-\phi_{0}(n-1))^{2}\right)
-i(q(n)\phi_{0}(n-1)-\phi_{0}(n)q(n-1))\non\\
&&-i\left\{2(\phi_{{\rm F}}-\phi_{{\rm I}})L/\lambda+(q_{{\rm F}}\phi_{{\rm F}}-q_{{\rm I}}\phi_{{\rm I}})\right\}\bigg\}
\non\\
&=&\sum_{n=1}^{N}\left\{-\frac{1}{2}(z^{*}(n)z(n)+z^{*}(n-1)z(n-1))+z^{*}(n)z(n-1)\right\}\non\\
&&-i\left\{2(\phi_{{\rm F}}-\phi_{{\rm I}})L/\lambda+(q_{{\rm F}}\phi_{{\rm F}}-q_{{\rm I}}\phi_{{\rm I}})\right\}. 
\label{red-scs-sc-z}
\eeqa
This expression, except for the last constant term, 
formally coincides with the corresponding action appearing in 
the (boson-)coherent-state path integral with 
a single degree of freedom. 
The last term can be neglected because 
it is a constant phase, which does not affect any physical quantity.

\subsection{Effective action for a domain wall in continuous-time formalism}

Let us comment on the continuous-time treatment of a domain wall 
and the associated problems.

If one started from (\ref{conti-ta}), 
one would obtain the continuous-time counterpart of (\ref{scs-s}) as 
\sbeqa
\beqa
&&{\cal S}^{{\rm s}}_{{\rm con}}[z^{*},z]
=\int_{0}^{T}dt{\cal L}_{{\rm con}},\non\\
&&{\cal L}_{{\rm con}}:=\frac{\hbar}{i}\frac{N_{{\rm DW}}S}{2}
(\dot{z}^{*}(t)z(t)-z^{*}(t)\dot{z}(t))
-E_{{\rm DW}}
\left[ 1+\frac{\alpha}{4}\left\{1+\cosh(z^{*}(t)-z(t))\right\}\right]\non\\
&& = 2\frac{N_{{\rm DW}}\hbar S}{\lambda}\dot{Q}(t)\phi_{0}(t)
-2N_{{\rm DW}}\hbar S\Omega\cos^{2}\phi_{0}(t)
\label{conti-Lag},\\
&&\Omega \equiv \frac{K}{2\hbar}\left(S-\frac{1}{2}\right)\alpha.
\eeqa
\seeqa 
In the last expression, constant terms including those appearing as a 
result of partial integration have been neglected. 
If the transverse anisotropy is strong in the sense that $N_{{\rm DW}}S\Omega\gg1$, 
$\phi_{0}(t)$ may be restricted to a region near $\pm \pi/2$; 
\beqa
\phi_{0}(t) = C\pi/2 + \varphi(t),\qquad 
C\equiv \pm 1, \qquad |\varphi(t)| \ll 1.
\eeqa
Substituting this into (\ref{conti-Lag}), one would find
\beqa
{\cal L}_{{\rm con}} \simeq CA\dot{Q}(t) 
+ \frac{2}{\pi}A\dot{Q}(t)\varphi(t) 
-2N_{{\rm DW}}\hbar S \Omega \varphi^{2}(t),
\label{conti-Lag2}
\eeqa
where $A\equiv N_{{\rm DW}}\hbar S \pi/\lambda$. 
Provided that the path-integration measure is independent of $\varphi$, 
Gaussian integration with respect to $\varphi$ would then 
lead to the following effective action 
for $Q$:
\beqa
\frac{i}{\hbar}{\cal S}^{{\rm eff}}_{{\rm con}}[Q] = \frac{i}{\hbar}\int_{0}^{T}dt \left\{CA\dot{Q}(t) + \frac{M_{{\rm D}}}{2}\dot{Q}^{2}(t)\right\},
\label{conti-action2}
\eeqa
where $M_{{\rm D}}$ is the D${\rm \ddot{o}}$ring mass:
\beqa
M_{{\rm D}} \equiv \frac{2\hbar^{2}}{a^{2}}\sqrt{\frac{KS}{\tilde{J}(S-1/2)}}\frac{1}{K_{\bot}}.
\eeqa
Accordingly, 
one might expect the following expression for the transition amplitude:
\beqa
\<z_{{\rm F}}|e^{-i\hat{H}T/\hbar}|z_{{\rm I}}\> \sim
\int {\cal D}Q \exp\left(\frac{i}{\hbar}{\cal S}^{{\rm eff}}_{{\rm con}}[Q]\right).
\label{conti-ta2}
\eeqa
This corresponds to the result of Braun and Loss.\cite{Braun-Loss}
At this stage, it has been concluded that 
the center position of the domain wall 
behaves as a free particle with a possible modification due to 
interference effects induced by the first term $CA\dot{Q}(t)$ 
of the effective action, which is often called the "Berry-phase term".

If one is interested in the quantum depinning of the domain wall, 
a pinning potential is to be added to Lagrangian (\ref{conti-Lag2}). 
If the effect of the transverse anisotropy is  relatively larger than 
that of the pinning potential, the effective action (\ref{conti-action2}) 
is simply augmented by the pinning potential. 
MQT of the center position has been discussed on the basis of this action. 
On the other hand, if the pinning effect is the larger, 
one could carry out Gaussian integration with respect to $Q(t)$ 
and obtain an effective action governing $\phi_{0}(t)$, with which 
MQC of the chirality has been discussed.\cite{Takagi-Tatara}

However, the whole series of the above-quoted arguments, which are 
based on the form of the derived effective actions 
regardless of the path-integration measure, are at best heuristic and 
their validity is rather dubious. 
In the literature it has been tacitly assumed that the right-hand side 
of (\ref{conti-ta2}) is a Feynman kernel with respect to $Q$. 
(Recall the following point: 
In order to obtain a transition amplitude 
between physical (i.e., normalizable) states from a Feynman kernel, 
the latter has to be multiplied by 
the initial and the final wave functions 
and integrated over $Q_{{\rm F}}$ and $Q_{{\rm I}}$.) 
The starting point of the whole arguments, on the other hand, is 
the left-hand side of (\ref{conti-ta2}), which is a transition amplitude 
between physical states. 
Its path-integral structure is different from that for a Feynman kernel. 
One should not be misled by the apparent form of the effective action 
(\ref{conti-action2}). 
Though integration over $\varphi(t)$ may be carried out in principle, 
resulting action can not be a Feynman kernel. 
Indeed, as we shall illustrate in Sec.VI, the 
right-hand side of (\ref{conti-ta2}) 
as interpreted as a Feynman kernel 
does not give a correct transition amplitude. 
By the same token, interference effects predicted on the basis of 
the "Berry-phase term" $CA\dot{Q}(t)$ need be re-examined: 
as shown in the following section, since the continuous-time 
treatment neglects 
many other terms which can contribute to interference effects, 
there is no reason that only the "Berry-phase term" should be kept.

\section{Transition Amplitude in Stationary-Action Approximation }
In this section, we evaluate transition amplitudes by means of 
the stationary-action approximation, 
discuss the conjugate relation between the center position and the chirality 
of the domain wall, and locate those terms which can induce interference effects.
\subsection{Stationary-action path}

Let $\{ \bzs(n),\zs(n) \}$ be the stationary-action path, namely, the stationary point of the action ${\cal S}^{{\rm s}}[z^{*},z]$:
\sbeqa
\beqa
&&\left.\frac{\partial {\cal S}^{{\rm s}}[z^{*},z]}
{\partial z^{*}(n)}\right|_{{\rm s}} = 0,\qquad  n=1,...,N-1, \\
&& \left.\frac{\partial {\cal S}^{{\rm s}}[z^{*},z]}
{\partial z(n-1)}\right|_{{\rm s}} = 0 ,\qquad  n=2,...,N,
\eeqa
\seeqa 
where the symbol $|_{{\rm s}}$ indicates the replacement $(z^{*},z)$ $\to$ $(\bzs,\zs)$ after differentiation. 
It is convenient to define\cite{Shibata-Takagi}
\beqa
\zs(0) := z_{{\rm I}} = q_{{\rm I}}+i\phi_{{\rm I}},\qquad \bzs(N) := z^{*}_{{\rm F}} = q_{{\rm F}}-i\phi_{{\rm F}}.
\label{bc1}
\eeqa
Let us work with (\ref{red-scs-sc-z}) instead of (\ref{red-scs-s}). 
This procedure will be justified {\it a posteriori}.
Then, the above set of equations take the form:
\sbeqa
\label{total-colle-eqmo}
\beqa
\zs(n)-\zs(n-1) &=& -i\epsilon\Omega \sinh \{ \bzs(n)-\zs(n-1)\},\qquad n=1,...,N-1,
\label{total-coll-eqmo1}\\
\bzs(n)-\bzs(n-1) &=& -i\epsilon\Omega \sinh \{ \bzs(n)-\zs(n-1)\},\qquad n=2,...,N.
\label{total-colle-eqmo2}
\eeqa
\seeqa 
By use of $\zs(n)$ and $\bzs(n)$, we define the stationary-action path for 
the center position and the chirality as 
\sbeqa
\label{qp-zbz}
\beqa
q^{{\rm s}}(n) &:=& (\zs(n)+\bzs(n))/2,
\label{qs-zsbzs}\\
\phi^{{\rm s}}(n) &:=& (\zs(n)-\bzs(n))/2i,
\label{ps-zsbzs}
\eeqa
\seeqa 
which are not real in general. 

The left-hand side of (\ref{total-colle-eqmo}) are $\order(\epsilon)$ 
because of the factor $\epsilon$ on the right-hand side. 
Hence, {\it as far as the equations for the stationary-action path 
is concerned, we can go over to the continuous time}:
\sbeqa
\label{conti-eqmo}
\beqa
\frac{d\zs(t)}{dt} &=& -i \Omega\sinh ( \bzs(t)-\zs(t) ),
\label{conti-eqmo1}\\
\frac{d\bzs(t)}{dt} &=& -i \Omega \sinh ( \bzs(t)-\zs(t) ).\label{conti-eqmo2}
\eeqa
\seeqa 
Note that the boundary condition is 
dictated by (\ref{bc1}) as 
\beqa
\zs(0)=z_{{\rm I}} ,\qquad \bzs(T)=z^{*}_{{\rm F}}.
\label{bc-conti}
\eeqa
It follows from Eqs. (\ref{conti-eqmo}) that 
\beqa
\frac{d}{dt}(\bzs(t)-\zs(t))= 0.
\eeqa
Hence, 
\beqa
\bzs(t)-\zs(t) = -2i\phi,
\label{sa}
\eeqa
where $\phi$ is a complex constant $(\phi \equiv \phi'+i\phi''; \phi',\phi'' \in {\bf R})$. 
Substituting (\ref{sa}) into (\ref{conti-eqmo}), and 
taking account of the boundary condition (\ref{bc-conti}), 
we obtain
\beqa
\zs(t) = -\Omega t \sin 2\phi +z_{{\rm I}}, 
\qquad 
\bzs(t) = -\Omega (t-T) \sin 2\phi + z^{*}_{{\rm F}}.
\label{sap}
\eeqa
Putting this back into (\ref{sa}), we find that the constant $\phi$ is 
determined by the following algebraic equation:
\beqa
\Omega T \sin 2\phi +  z^{*}_{{\rm F}} - z_{{\rm I}} = -2i\phi, 
\eeqa
or equivalently, 
\sbeqa
\label{alge-eq}
\beqa
&& 2\phi'' - \Omega T \sin 2\phi'\cosh 2\phi''= q \equiv q_{{\rm F}}-q_{{\rm I}},
\label{alge-eq1}\\
&& 2\phi' + \Omega T \cos 2\phi'\sinh 2\phi'' = \phi_{{\rm F}}+\phi_{{\rm I}}.
\label{alge-eq2}
\eeqa
\seeqa

The stationary-action path can be expressed in terms of 
$q^{{\rm s}}(t)$ and $\phi^{{\rm s}}(t)$ as defined by (\ref{qp-zbz}) as 
\sbeqa
\label{sap-qp}
\beqa
q^{{\rm s}}(t) &=& -\Omega t \sin 2\phi + z_{{\rm I}}-i\phi,
\label{sap-qp1}\\
\phi^{{\rm s}}(t) &=& \phi.
\label{sap-qp2}
\eeqa
\seeqa 
Note that both of these are complex. 
Eq. (\ref{sap-qp2}) shows that $\phi^{{\rm s}}(n)-\phi^{{\rm s}}(n-1) = 0 $ 
for all $n$ . 
This justifies our procedure of working with (\ref{red-scs-sc-z}) instead of (\ref{red-scs-s}). 
Incidentally it follows from (\ref{sap-qp}) that 
\beqa
\frac{d q^{{\rm s}}(t)}{dt} = -\Omega\sin 2\phi^{{\rm s}}(t).
\label{dq-p}
\eeqa
Thus, the velocity of the center position depends on the chirality and 
is proportional to the transverse anisotropy 
(recall that $\Omega\propto\alpha$). 
In the special circumstance that $\phi^{{\rm s}}(t)$ happens to be close to 
$\pm\pi/2$, we can put $\phi^{{\rm s}}(t) = \pm \pi/2 + \varphi^{{\rm s}}(t)$ 
to find 
\beqa
\frac{d q^{{\rm s}}(t)}{dt} \propto \varphi^{{\rm s}}(t).
\eeqa
This shows that in such a circumstance the center position 
and the chirality are mutually canonically conjugate, thereby confirming the 
claim made in Ref. 10. 
In any case it is clear that $q^{{\rm s}}(t)$ and $\phi^{{\rm s}}(t)$ are 
closely coupled; 
they should be treated on an equal footing.

\subsection{Stationary action}

The stationary action ${\cal S}^{{\rm s}}[\bzs,\zs]$ 
may be arranged as 
\sbeqa
\label{sa-ss}
\beqa
{\cal S}^{{\rm ss}} &:=&{\cal S}^{{\rm s}}[\bzs,\zs]
={\cal S}^{{\rm ssc}}+{\cal S}^{{\rm ssd}},\\
\frac{i}{\hbar}{\cal S}^{{\rm ssc}}&:=&
N_{{\rm DW}}S\Bigg[-\frac{1}{2}(|z_{{\rm F}}|^{2}+|z_{{\rm I}}|^{2})+\frac{1}{2}(z^{*}_{{\rm F}}\zs(N-1)+\bzs(1)z_{{\rm I}})\non\\
&&+\frac{1}{2}\sum_{n=1}^{N-1}\{
(\bzs(n+1)-\bzs(n))\zs(n)-\bzs(n)(\zs(n)-\zs(n-1))
\}\Bigg],\label{sa-ssc}\\
\frac{i}{\hbar}{\cal S}^{{\rm ssd}}&:=& -\frac{i}{\hbar}E_{{\rm DW}}\sum_{n=1}^{N}\epsilon
\Bigg[
1+\frac{\alpha}{4}\left\{1+\cosh(\bzs(n)-\zs(n-1))\right\}\Bigg].
\label{sa-ssd}
\eeqa
\seeqa
The first and the second terms on the right-hand side of (\ref{sa-ssc}) 
depend on the initial and the final state. 
The second term depends also on $T$  through the stationary-action path. 
These terms, which have been neglected in the continuous-time formalism, 
turn out to be crucial for a correct evaluation of the transition amplitude. 
The third term corresponds to the "Berry-phase term" 
in the continuous-time formalism. 
In the latter formalism, interference effects have been ascribed to this 
term alone. 
However, the second term can also contribute to interference effects. 
This is another remarkable difference from the continuous-time formalism. 
Of course, interference effects in question can arise only if 
there exist two or more stationary-action paths. 
In the case of a free domain wall under consideration, 
there is no question of interference 
because the stationary-action path is unique.

Substituting (\ref{sap}) into (\ref{sa-ss}), we obtain
\sbeqa
\beqa
&&\frac{i}{\hbar}{\cal S}^{{\rm sc}} = 
-\frac{N_{{\rm DW}}S}{2}\left[|z_{{\rm F}}|^{2}+|z_{{\rm I}}|^{2}-2z^{*}_{{\rm F}}z_{{\rm I}}
-(\Omega T \sin 2\phi)^{2}\right] +\order(\epsilon),\\
&&\frac{i}{\hbar}{\cal S}^{{\rm sd}} = 
-\frac{i}{\hbar}E_{{\rm DW}}T\left(1+\frac{\alpha}{4}\right)-iN_{{\rm DW}}S\Omega T
\cos 2\phi + \order(\epsilon).
\eeqa
\seeqa
Putting these together, 
we finally arrive at 
\beqa
\frac{i}{\hbar}{\cal S}^{{\rm ss}} &=& -\frac{N_{{\rm DW}}S}{2} \Bigg[
|z_{{\rm F}}|^{2}+|z_{{\rm I}}|^{2}-2z^{*}_{{\rm F}}z_{{\rm I}}-(\Omega T\sin 2\phi)^{2}+2i\Omega T\cos2\phi
\Bigg]\non\\
&&-\frac{i}{\hbar}E_{{\rm DW}}T\left(1+\frac{\alpha}{4}\right).
\eeqa
This stationary action is complex in general. 

\subsection{Fluctuations}
 
As noted previously,\cite{Shibata-Takagi} 
well-defined evaluation of the fluctuation integral is possible 
only in the discrete-time formalism. Accordingly 
we separate the integration variables as
\sbeqa
\beqa
&&z(n) = \zs(n)+ \zeta(n),
\qquad
z^{*}(n) = \bzs(n) + \zeta^{*}(n)~~:n=1,2,...,N-1.\label{fluc}
\eeqa
It is convenient to define
\beqa 
\zeta(0)=\zeta^{*}(N)=0. 
\eeqa
\seeqa
Substituting  (\ref{fluc}) into the action (\ref{scs-sd}) and 
(\ref{red-scs-sc-z}), and expanding up to the second order in the fluctuation, 
we get 
\sbeqa
\beqa
&& {\cal S}^{{\rm s}}[z^{*},z] = {\cal S}^{{\rm ss}} + {\cal S}_{2}[\zeta^{*},\zeta],\\
&& {\cal S}_{2}[\zeta^{*},\zeta] = 
{\cal S}^{{\rm c}}_{2}[\zeta^{*},\zeta]+{\cal S}^{{\rm d}}_{2}[\zeta^{*},\zeta],
\eeqa
\seeqa 
where
\sbeqa
\label{fluc-ss}
\beqa
&&\frac{i}{\hbar}{\cal S}^{{\rm c}}_{2}[\zeta^{*},\zeta]
:= -\frac{N_{{\rm DW}}S}{2}\sum_{n=1}^{N}
\left\{2\zeta^{*}(n)\zeta(n)-2\zeta^{*}(n)\zeta(n-1)\right\},
\label{fluc-ssc}\\
&& \frac{i}{\hbar}{\cal S}^{{\rm d}}_{2}[\zeta^{*},\zeta]
:= -i\frac{N_{{\rm DW}}S}{2}\Omega\cos2\phi\sum_{n=1}^{N}\epsilon
\left\{(\zeta^{*}(n))^{2} + (\zeta(n-1))^{2} -2\zeta^{*}(n)\zeta(n-1) \right\}.
\label{fluc-ssd}
\eeqa
\seeqa 
Accordingly, the transition amplitude reduces to
\sbeqa
\beqa
&&\<z_{{\rm F}}|e^{-i\hat{H}T/\hbar}|z_{{\rm I}}\>
\simeq \exp \left[\frac{i}{\hbar}{\cal S}^{{\rm ss}}\right]K_{2}(T),\label{final-ta}\\
&& K_{2}(T) := \lim_{N\to\infty} \int \prod_{n=1}^{N-1} \frac{1}{{\cal M}}
\frac{d\zeta(n)d\zeta^{*}(n)}{2\pi i}
\exp\left(\frac{i}{\hbar}{\cal S}_{2}[\zeta^{*},\zeta]\right),\label{K2}
\eeqa
\seeqa 
where ${\cal M}$ is a constant, whose value as well as 
the detailed evaluation of $K_{2}(T)$ are given in Appendix B. 
The result is 
\beqa
K_{2}(T) = \frac{e^{i\Omega_{\phi}T/2}}{\sqrt{1-i\Omega_{\phi}T}},
\qquad
\Omega_{\phi} \equiv \Omega \cos 2\phi. \label{K2T}
\eeqa
This completes a microscopic evaluation of transition amplitudes 
for a free domain wall. 

\section{Transition Probability}

We can now compute transition probabilities 
between various initial and final domain-wall states:
\sbeqa
\beqa
P(q,\phi_{{\rm F}},\phi_{{\rm I}};T)
&:=& |\<z_{{\rm F}}|e^{-i\hat{H}T/\hbar}|z_{{\rm I}}\>|^{2}\non\\
&\simeq& \frac{e^{\Im \Omega_{\phi} T}}{\sqrt{1+2\Im \Omega_{\phi}T +|\Omega_{\phi}|^{2}T^{2}}}\exp\left[-\frac{2}{\hbar}\Im {\cal S}^{{\rm ss}}\right],
\label{tp}\\
\frac{2}{\hbar}\Im {\cal S}^{{\rm ss}} &=& N_{{\rm DW}}S\bigg[ 
q^{2}+ (\phi_{{\rm F}}-\phi_{{\rm I}})^{2}
-2(\Omega T)^{2}\Re (\sin 2\phi)^{2}
-2\Omega T \Im \cos 2\phi
\bigg].
\label{real-action}
\eeqa
\seeqa
The right-hand side does not depend on $q_{{\rm F}}$ and $q_{{\rm I}}$ separately but 
only on $q(\equiv q_{{\rm F}}-q_{{\rm I}})$ as expected from the translation invariance. 
This justifies the notation $P(q, \phi_{{\rm F}}, \phi_{{\rm I}} ;T)$. 

\subsection{Case of $\alpha=0$}

In the absence of transverse anisotropy $(\Omega\propto\alpha = 0)$, 
the transition probability is independent of $T$:
\beqa
P(q,\phi_{{\rm F}},\phi_{{\rm I}};T)&=&\exp\left[-N_{{\rm DW}}S \left\{ q^{2}+(\phi_{{\rm F}}-\phi_{{\rm I}})^2 \right\}
\right]= |\<z_{{\rm F}}|z_{{\rm I}}\>|^{2}.\label{notrans-tp}
\eeqa
Thus, the transition probability 
coincides with the overlap between the initial and the final states 
and is a function of the 
differences $q_{{\rm F}}-q_{{\rm I}}$ and $\phi_{{\rm F}}-\phi_{{\rm I}}$. 
It is shown in Fig. 3.

The factor $N_{{\rm DW}}S$ in the exponent of (\ref{notrans-tp}) 
is large (typically of order of $10^{2}$), 
reflecting the semi-macroscopic character of a typical domain wall of interest. 
Hence, 
if the final state is even slightly different from the initial state, 
the transition is forbidden.

\subsection{Case of $\alpha \neq0$}

In the presence of the transverse anisotropy, 
the transition probability depends on $T$:
\beqa
&&P(q,\phi_{{\rm F}},\phi_{{\rm I}};T)\non\\
&&=\frac{\exp(-\tau \sin 2\phi'\sinh 2\phi'')}{\sqrt{
1-2\tau \sin 2\phi'\sinh 2\phi''+\tau^{2}
\{(\cos 2\phi'\cosh 2\phi'')^{2}+(\sin 2\phi'\sinh 2\phi'')^{2}
\}}}\non\\
&&\times \exp\Bigg[-N_{{\rm DW}}S\bigg\{
q^{2}+(\phi_{{\rm F}}-\phi_{{\rm I}})^{2}\non\\
&&-\tau^{2}
\bigg((\sin 2\phi'\cosh 2\phi'')^{2}-(\cos 2\phi'\sinh 2\phi'')^{2}\bigg)
+2\tau \sin 2\phi'\sinh 2\phi''
\bigg\}\Bigg],
\label{exact-tp}
\eeqa
where $\tau \equiv \Omega T$. 

\subsubsection{analytical evaluation in linear approximation}
In order to find $\phi(\equiv \phi'+i\phi'')$, 
we need to solve the algebraic equation (\ref{alge-eq}).
On inspection we see that it has a solution 
$\phi=(\phi_{{\rm F}}+\phi_{{\rm I}})/2$ if 
$q=-\tau\sin(\phi_{{\rm F}}+\phi_{{\rm I}})$. 
This motivates us to look for a more general class of solutions 
by linearizing (\ref{alge-eq}) under 
the following condition to be justified {\it a posteriori}:
\beqa
\phi' = \frac{\phi_{{\rm F}}+\phi_{{\rm I}}}{2} + \vap',
\qquad \phi'' = \vap'',
\qquad |\vap'|,|\vap''|\ll 1.
\label{asump}
\eeqa
Then, we can write down the linearized version of (\ref{alge-eq}) as 
\beqa
2\vap''-\tau C_{{\rm FI}}2\vap' =q',
\qquad 2\vap' + \tau C_{{\rm FI}}2\vap'' = 0,
\eeqa
where 
\beqa
q'\equiv q+\tau S_{{\rm FI}},
\qquad C_{{\rm FI}} \equiv \cos(\phi_{{\rm F}}+\phi_{{\rm I}}), 
\qquad S_{{\rm FI}} \equiv \sin(\phi_{{\rm F}}+\phi_{{\rm I}}).
\eeqa 
Hence
\beqa
2\vap' = -\frac{\tau C_{{\rm FI}}}{1 +(\tau C_{{\rm FI}})^{2}}q',\qquad 2\vap'' =\frac{1}{1+(\tau C_{{\rm FI}})^{2}}q'.
\label{vap-q}
\eeqa
Since the factors multiplying $q'$ are at most of order unity, 
the assumed condition (\ref{asump}) is satisfied if $|q'|\ll 1$. 
Substituting (\ref{vap-q}) into (\ref{real-action}) 
we find  
\beqa
-\frac{2}{\hbar}\Im {\cal S}^{{\rm ss}} = -N_{{\rm DW}}S\left[
\frac{q'^{2}}{1+(\tau C_{{\rm FI}})^{2}} + (\phi_{{\rm F}}-\phi_{{\rm I}})^{2}\right],
\label{li-sa}
\eeqa
while the prefactor of (\ref{tp}) is obtained, up to $\order(q'^{2})$, as
\beqa
\frac{1}{\sqrt{1+(\tau C_{{\rm FI}})^{2}}}\left[
1-\left(\frac{\tau C_{{\rm FI}}}{1+(\tau C_{{\rm FI}})^{2}}\right)^{2}
\left\{
2\tau S_{{\rm FI}}q'+
\left(\frac{1}{2}\left(\frac{S_{{\rm FI}}}{C_{{\rm FI}}}\right)^{2}
-\frac{(\tau C_{{\rm FI}})^{2}}{1+(\tau C_{{\rm FI}})^{2}}
\right)q'^{2}
\right\}
\right].
\eeqa
Because of the large factor $N_{{\rm DW}}S$ in the exponent of (\ref{li-sa}), 
the $q'$-dependence of the prefactor is negligible. 
Hence, 
\beqa
P(q,\phi_{{\rm F}},\phi_{{\rm I}};T)
&\simeq&\frac{1}{\sqrt{1+(\tau \cos(\phi_{{\rm F}}+\phi_{{\rm I}}))^{2}}}\non\\
&&\times 
\exp\left[-N_{{\rm DW}}S\left\{
\frac{(q+\tau \sin(\phi_{{\rm F}}+\phi_{{\rm I}}))^{2}}{1+(\tau \cos(\phi_{{\rm F}}+\phi_{{\rm I}}))^{2}}+(\phi_{{\rm F}}-\phi_{{\rm I}})^{2}
\right\}
\right],\label{li-tp}
\eeqa
which is valid provided that $|q+\tau \sin(\phi_{{\rm F}}+\phi_{{\rm I}})|\ll 1$. 
This approximate formula 
shows that transitions are suppressed if $\phi_{{\rm F}}\neq\phi_{{\rm I}}$. 
It also suggests the following picture: 
for a given $(\phi_{{\rm F}},\phi_{{\rm I}})$, the peak of the wave packet 
representing the center position $q$ moves with the velocity 
(in units of $\lambda\Omega$) 
\beqa
v_{{\rm packet}} \equiv -\sin(\phi_{{\rm F}}+\phi_{{\rm I}}),
\label{velocity}
\eeqa
while the width (in units of $\lambda$) of the wave packet increases as 
\beqa
w_{{\rm packet}}(\tau) \equiv 
\{1+(\tau\cos(\phi_{{\rm F}}+\phi_{{\rm I}}))^{2}\}^{1/2}. 
\label{width}
\eeqa
The minimal velocity and maximal spreading occurs at 
$|\phi_{{\rm F}}+\phi_{{\rm I}}|=0~ ({\rm mod}~\pi)$, 
while the maximal velocity and minimal spreading occurs at 
$|\phi_{{\rm F}}+\phi_{{\rm I}}|=\pi/2~ ({\rm mod}~\pi)$. 
We choose these cases as well as the intermediate case of
$|\phi_{{\rm F}}+\phi_{{\rm I}}|=2\pi/3$, and depict (\ref{li-tp}) 
in Figs. 4$-$6 with solid curves. These figures show the case of 
minimal suppression ($\phi_{{\rm F}}=\phi_{{\rm I}}$). 
We take $N_{{\rm DW}}S=100$ throughout (and also 
in Fig. 7 to be mentioned below).

Fig. 4 shows the case of $\phi_{{\rm F}}=\phi_{{\rm I}}=\pi/2$, where 
$v_{{\rm packet}}=0$ and $w_{{\rm packet}}(\tau)=(1+\tau^{2})^{1/2}$. 
The curve for $q=0$, 
where the formula (\ref{li-tp}) is exact as mentioned at the beginning 
of this subsection, 
shows that the probability of remaining in the initial state 
decreases with time. 
Thus the domain wall exhibits a quantum phenomenon analogous to the 
{\it wave-packet spreading} for a free particle. 
The long-time tail originates from the prefactor of 
(\ref{exact-tp}) coming from the fluctuation. 
If the final state corresponds to the mere displacement of 
the initial domain wall by the distance $q\neq 0$, 
the transition probability exhibits an initial increase, 
which may be interpreted as the
appearance of an overlap between the final wave packet and the 
wave packet evolving from the initial state. 
For a small $q$, the overlap initially suppressed by the 
large factor $N_{{\rm DW}}S$ is rapidly recovered resulting in a 
sharp initial increase of the transition probability. 
The origin of the long-time tail is the same in the case of $q=0$. 
The case of $\phi_{{\rm F}}=\phi_{{\rm I}}=-\pi/3$, where 
$v_{{\rm packet}}=\sqrt{3}/2$ and 
$w_{{\rm packet}}(\tau)=(1+\tau^{2}/4)^{1/2}$, 
is shown in Fig. 5, which exhibits a moving and spreading 
wave packet. So far, the quantum dynamics of the domain wall 
resembles that of a free particle. 
However, Fig. 6, where $\phi_{{\rm F}}=\phi_{{\rm I}}=-\pi/4$, 
reveals quite a different feature; 
it may be interpreted as showing a non-spreading wave packet 
like a solitary wave with 
$v_{{\rm packet}}=1$ and $w_{{\rm packet}}(\tau)=1$. 
Finally, 
Fig. 7 shows that the transition probabilities 
between states with the common center position ($q=0$) and 
different chiralities are rather small. 
This is in contrast to the situation in Fig. 4, 
where there is an appreciable transition probability 
between states with the common chirality ($\phi_{{\rm F}}=\phi_{{\rm I}}$) 
and $q\neq 0$.

\subsubsection{numerical evaluation}
To check the accuracy of the approximate formula (\ref{li-tp}), 
we have numerically solved (\ref{alge-eq}) for $\phi'$ and $\phi''$, 
and put them into (\ref{exact-tp}). 
The results are depicted in Figs. 4$-$7. 
It is confirmed that (\ref{li-tp}) remains valid even for 
$|q+\tau\sin(\phi_{{\rm F}}+\phi_{{\rm I}})|\sim 1$.

\subsection{Comparison with quantum dynamics of a free particle}

Let us compare the above-found quantum behavior of the domain wall 
with that of a free particle. 

The transition amplitude for a free particle with mass $m$ 
between the initial coherent state $|Z_{{\rm I}}\>(Z_{{\rm I}}=Q_{{\rm I}}+iP_{{\rm I}})$ 
and the final coherent state $|Z_{{\rm F}}\>(Z_{{\rm F}}=Q_{{\rm F}}+iP_{{\rm F}})$ is given by
\sbeqa
\beqa
\<Z_{{\rm F}}|e^{-i\hat{H}T/\hbar}|Z_{{\rm I}}\>&=&\lim_{N\to\infty}\int\prod_{n=1}^{N-1}
\frac{dZ(n)dZ^{*}(n)}{2\pi i}
\exp\left(\frac{i}{\hbar}{\cal S}[Z^{*},Z]\right),\\
\frac{i}{\hbar}{\cal S}[Z^{*},Z]
&:=&\sum_{n=1}^{N}\Bigg[
-\frac{1}{2}(|Z(n)|^{2}+|Z(n-1)|^{2}+Z^{*}(n)Z(n-1)\non\\
&&-i\frac{\epsilon}{8}
\{1-(Z^{*}(n)-Z(n-1))^{2}\}
\Bigg].
\label{cs-action}
\eeqa
\seeqa
This action resembles that of a free domain wall 
(\ref{scs-sd}) and (\ref{red-scs-sc-z}). 
Indeed, if we expand the non-linear part in (\ref{scs-sd}) as
\beqa
\cosh(z^{*}(n)-z(n-1)) \simeq 1-\frac{(z^{*}(n)-z(n-1))^2}{2},
\eeqa
then the action (\ref{scs-s}) for a free domain wall 
reduces to (\ref{cs-action}). 
In this sense, a free domain wall may be said to be a non-linear version of 
a free particle.

The transition probability for the free particle is exactly calculated as
\beqa
&&|\<Z_{{\rm F}}|e^{-i\hat{H}T/\hbar}|Z_{{\rm I}}\>|^{2}=
\left({1+\left(\frac{\hbar T}{2m\delta^{2}}\right)^{2}}\right)^{-1/2}\non\\
&&\times
\exp\left[
-\frac{1}{2\delta^{2}}
\frac{1}{1+\left(\frac{\hbar T}{2m\delta^{2}} \right)^{2}}\left\{
 (Q_{{\rm F}}-Q_{{\rm I}})-\frac{T}{2m}(P_{{\rm F}}+P_{{\rm I}})\right\}^{2}
-\frac{\delta^{2}}{2\hbar^{2}}\left(P_{{\rm F}}-P_{{\rm I}}\right)^{2}
\right],
\label{tp-one2}
\eeqa
where $Q_{{\rm I}}$ and $Q_{{\rm F}}$ are the center of 
the initial and the final Gaussian wave packet, respectively, 
and likewise $P_{{\rm I}}$ and $P_{{\rm F}}$ are the initial and the final mean
momentum. 
The width of the wave packet has been chosen to be $\delta$ both for 
the initial and the final state. 
The form of (\ref{li-tp}) differs from that of (\ref{tp-one2}) 
only in the non-linear factors $\cos(\phi_{{\rm F}}+\phi_{{\rm I}})$ and 
$\sin(\phi_{{\rm F}}+\phi_{{\rm I}})$.
If these are linearized as 
$\cos(\phi_{{\rm F}}+\phi_{{\rm I}})\simeq 1$ and 
$\sin(\phi_{{\rm F}}+\phi_{{\rm I}}) \simeq \phi_{{\rm F}}+\phi_{{\rm I}}$, 
then 
(\ref{li-tp}) reduces to the same form as (\ref{tp-one2}); 
in this case the behavior of a free domain wall 
is the same as that of a free particle. 
However, the linearization is not always allowed. 
In particular, 
when $\phi_{{\rm F}}+\phi_{{\rm I}}=\pm \pi/2$, 
the behavior of the domain wall is completely different from 
that of the free particle; 
the wave packet does not spread!
This is a manifestation of the non-linear character of spin.

\subsection{Effective mass of a free domain wall}
We can estimate the effective domain-wall mass 
from the correspondence of the domain wall and 
the particle as noted in the previous subsection. 

Comparison of (\ref{li-tp}) and (\ref{tp-one2}) reveals 
the following  correspondence. 
First, the coefficient of $T$ in the prefactor suggests
\beqa
\Omega|\cos(\phi_{{\rm F}}+\phi_{{\rm I}})| \longleftrightarrow  \frac{\hbar}{2m\delta^{2}}.
\eeqa
Second, the coefficient of $q^{2}$ in the exponent at $T=0$ suggests
\beqa
\delta_{{\rm DW}} := \frac{\lambda}{\sqrt{2N_{{\rm DW}}S}} \longleftrightarrow \delta.
\eeqa
$\delta_{{\rm DW}}$ can thus be interpreted as the width of the wave packet 
describing the initial domain wall. 
These two correspondences suggest 
to associate the domain wall with 
the "effective mass" $M_{{\rm DW}}$ given by
\beqa
M_{{\rm DW}}= \frac{\hbar}{2\Omega|\cos(\phi_{{\rm F}}+\phi_{{\rm I}})|\delta^{2}_{{\rm DW}}} 
      = \frac{M_{{\rm D}}}{|\cos(\phi_{{\rm F}}+\phi_{{\rm I}})|}.
\eeqa
It is to be noted, however, that this "effective mass" depends on the 
initial and the final chirality; 
as such, it can not be viewed as an effective mass 
of  an ordinary dynamical entity. 
It coincides with the D${\rm \ddot{o}}$ring mass 
if $\phi_{{\rm F}}+\phi_{{\rm I}} = 0$ or $\pm\pi$. 
On the other hand, 
it is infinite if $\phi_{{\rm F}}+\phi_{{\rm I}} =\pm \pi/2$, 
which is just another way of expressing the 
non-spreading of the wave packet as noted 
in the previous subsection.

\section{Discussion}


If we calculated the transition probability by use of the
continuous-time action (\ref{conti-Lag}), 
what result would have been obtained? 
Though the stationary-action path is the same as (\ref{sap}), 
the stationary action would be 
\beqa
\frac{i}{\hbar}{\cal S}^{{\rm ss}}_{{\rm con}}&:=& 
\frac{i}{\hbar}{\cal S}^{{\rm s}}_{{\rm con}}[\bzs,\zs]\non\\
&=& -iN_{{\rm DW}}S\tau(\phi\sin2\phi+\cos2\phi)-\frac{i}{\hbar}E_{{\rm DW}}T
\left(1+\frac{\alpha}{4}\right).
\eeqa
Hence
\beqa
&&|\<z_{{\rm F}}|e^{-i\hat{H}T/\hbar}|z_{{\rm I}}\>|^{2}
\sim |K_{2{\rm con}}(T)|^{2}\exp\left(-\frac{2}{\hbar}
\Im {\cal S}^{{\rm ss}}_{{\rm con}}\right), \\
&&-\frac{2}{\hbar}\Im{\cal S}_{{\rm con}}^{ss}=
N_{{\rm DW}}S\tau (
\phi''\sin 2\phi'\cosh 2\phi''+\phi'\cos 2\phi' \sinh 2\phi''-\sin 2\phi'\sinh 2\phi'' 
).
\eeqa
Let $T=0$. 
Assuming that the fluctuation integral $K_{2{\rm con}}(T)$ was somehow 
evaluated and that it agreed with the correct value (\ref{K2T}), 
we would have found 
$|\<z_{{\rm F}}|e^{-i\hat{H}T/\hbar}|z_{{\rm I}}\>|^{2}_{T=0}\sim 1$. 
However, this is a completely meaningless result,  
since the left-hand side should be equal to 
$|\<z_{{\rm F}}|z_{{\rm I}}\>|^{2}$. 
(If one started from the effective action (\ref{conti-action2}), 
one would also find a similarly meaningless result.)
Furthermore one can not 
rationally calculate the fluctuation integral 
from this formalism.\cite{Shibata-Takagi}

\section{Concluding Remarks}

We have considered the macroscopic quantum dynamics of a free 
domain wall in a quasi-one-dimensional ferromagnet by use of 
the spin-coherent-state path integral in the discrete-time formalism. 
The center position and the chirality, which have been chosen as the 
collective degrees of freedom, are noted to be mutually canonically conjugate 
in a loose sense. 
The quantum behavior of the domain wall 
is the same as that of the free particle and its effective mass is 
the D${\rm \ddot{o}}$ring mass if $\phi_{{\rm F}}+\phi_{{\rm I}}=0$ or $\pm \pi$, 
but in general it differs from the latter in some non-linear effects. 
We have also pointed out some grave difficulties associated with the 
continuous-time formalism. 
It can not correctly evaluate transition amplitudes. 
Its assertion on 
interference effects on the basis of the "Berry-phase term" 
alone is also questionable.

Let us speculate on MQT and MQC involving a domain wall. 
Since a free domain wall with a fixed chirality 
$(\phi_{{\rm F}}=\phi_{{\rm I}})$
has been shown to behave roughly like a free particle unless 
$\phi_{{\rm F}} = \phi_{{\rm I}} = \pm \pi/4$, we expect that 
a quantum depinning (MQT) will occur in the case of a weak pinning and 
a strong transverse anisotropy as mentioned by many workers; 
a strong transverse anisotropy tends to fix the chirality at 
$\phi_{{\rm F}}=\phi_{{\rm I}}=\pm \pi/2$. 
However, if a transverse anisotropy energy is 
comparable to a pinning potential, 
the dependence of the domain-wall mass on the chirality 
can be important. This may somewhat affect the MQT. 
Such a possibility has been overlooked in the literature. 
The MQC has been suggested to occur in the case of 
a strong pinning and a weak transverse anisotropy,
\cite{Takagi-Tatara} namely for a fixed center position 
$(q=0)$. 
However, Fig. 7 shows that the transition probability 
for $q=0$ and $|\phi_{{\rm F}}-\phi_{{\rm I}}| \sim \pi$ is negligible. 
This is due to the large factor $N_{{\rm DW}}S$ in the exponent. 
Hence, for the MQC to occur, it may be necessary to 
invoke a mechanism (e.g., a magnetic field \cite{Braun-Loss}) to 
decrease $|\phi_{{\rm F}}-\phi_{{\rm I}}|$. 
At any rate, a careful consideration is needed to 
make a conclusion on the possibility of the MQC.


\acknowledgments

We are grateful to T. Nakamura for many useful discussions.


\appendix
\section{Derivation of Eq. (3.9)}

The numerator in the logarithm of (\ref{scs-sc}) can be 
rewritten in two forms;
\sbeqa
\label{deform}
\beqa
&&1+e^{-x+z^{*}(n)+z(n-1)}\non\\
&&=(1+e^{-x+z^{*}(n)+z(n)})\left(1+
\frac{e^{-x+z^{*}(n)+z(n)}}{1+e^{-x+z^{*}(n)+z(n)}}(e^{-(z(n)-z(n-1))}-1)\right),\\
&&=(1+e^{-x+z^{*}(n-1)+z(n-1)})\left(1+
\frac{e^{-x+z^{*}(n-1)+z(n-1)}}{1+e^{-x+z^{*}(n-1)+z(n-1)}}(e^{(z^{*}(n)-z^{*}(n-1))}-1)\right).
\eeqa
\seeqa
In order to perform the $x$ integration, 
we pay attention to the following formula;
\beqa
\int dx \ln\left(1+\frac{A}{1+e^{x+B}}\right)&=&
Di\left(1+\frac{A}{1+e^{x+B}}\right)
-Di\left(\frac{1+\frac{A}{1+e^{x+B}}}{1+A}\right)\non\\
&&+\ln(1+A)\ln\left(\frac{-Ae^{x+B}}{1+e^{x+B}}\right),
\eeqa
where $Di(z)$ is the {\it dilogarithm} defined by \cite{Abramowitz-Stegun}
\beqa
Di(z) := \int_{1}^{z}dt\frac{\ln t}{1-t}.\label{Di}
\eeqa
Thus, 
the integral in (\ref{scs-sc}) 
can be evaluated as 
\beqa
{\it I} &\equiv&\int_{-L/\lambda}^{L/\lambda}dx \ln 
\frac{(1+e^{-x+z^{*}(n)+z(n-1)})^2}{(1+e^{-x+z^{*}(n)+z(n)})(1+e^{-x+z^{*}(n-1)+z(n-1)})}\non\\
&=&\sum_{\alpha=1}^{2}\Bigg[
Di\left(1+\frac{A_{\alpha}}{1+e^{x+B_{\alpha}}}\right)
-Di\left(\frac{1+\frac{A_{\alpha}}{1+e^{x+B_{\alpha}}}}{1+A_{\alpha}}\right)
\non\\
&+&\ln(1+A_{\alpha})\ln\left(\frac{-A_{\alpha}e^{x+B_{\alpha}}}{1+e^{x+B_{\alpha}}}\right)\Bigg]_{-L/\lambda}^{L/\lambda},
\label{int1}
\eeqa
where
\sbeqa
\label{AB}
\beqa
&&A_{1}\equiv e^{-(z(n)-z(n-1))}-1=e^{-(q(n)-q(n-1))-i(\phi_{0}(n)-\phi_{0}(n-1))}-1,\\
&&B_{1}\equiv -(z^{*}(n)-z(n))=-2q(n),\\
&&A_{2}\equiv e^{z^{*}(n)-z^{*}(n-1)}-1=e^{q(n)-q(n-1)-i(\phi_{0}(n)-\phi_{0}(n-1))}-1,\\
&&B_{2}\equiv -(z^{*}(n-1)-z(n-1))=-2q(n-1).
\eeqa
\seeqa
Since $L/\lambda \gg 1$, (\ref{int1}) may be simplified as 
\beqa
{\it I} = \sum_{\alpha=1}^{2}\Bigg[
-\left\{Di(1+A_{\alpha})+Di\left(\frac{1}{1+A_{\alpha}}\right)\right\}
+\ln(1+A_{\alpha})\left(\frac{L}{\lambda}+\ln\exp(-B_{\alpha})\right)
\Bigg].
\label{int2}
\eeqa
Next, we use the dilogarithm identity
\beqa
Di(A)+Di\left(\frac{1}{A}\right)= -\frac{1}{2}(\ln A)^{2}
\eeqa
to find 
\beqa
I &=& \sum_{\alpha=1}^{2}
\Bigg[
\frac{1}{2}(\ln(1+A_{\alpha}))^{2}+ \ln(1+A_{\alpha})
\left(\frac{L}{\lambda}+\ln\exp(-B_{\alpha})\right)
\Bigg] \non\\
&=&-\left[(q(n)-q(n-1))^{2}+R(\Delta\phi_{0}(n))
+i\left\{2(q(n)+q(n-1))+2L/\lambda\right\}I(\Delta\phi_{0}(n))\right],
\label{final-rep}
\eeqa
where
\sbeqa
\beqa
\Delta\phi_{0}(n)&\equiv& \phi_{0}(n)-\phi_{0}(n-1),\\
R(\phi)&:=&-\left(\ln\exp(-i\phi)\right)^{2} ~:|\phi|\le\pi,\label{peri-R}
\label{peri-R}\\
I(\phi)&:=&i\ln\exp(-i\phi)~~~~~~:-\pi\le\phi<\pi.
\label{peri-I}
\eeqa
\seeqa
$R(\phi)$ and $I(\phi)$, which are $2\pi$-periodic functions, 
can be cast into the form (\ref{periodic-func}). 
This periodicity follows from (\ref{scs-sc}), 
which originates from 
the overlap of between spin-coherent states (\ref{peri-c}) whose 
real and imaginary parts are even and odd periodic, respectively. 
The discontinuity or non-smoothness are 
the consequence of the spatial continuum approximation. 

\section{Calculation of Fluctuation Integral}
The fluctuation action (\ref{fluc-ss}) may be written as 
\beqa
\frac{i}{\hbar}{\cal S}^{{\rm ss}}_{2}(N-1) &:=& -i\frac{N_{{\rm DW}}S}{2}
\sum_{n=1}^{N-1}\left\{A(n)(\zeta^{*}(n))^{2}+B(n)(\zeta(n))^{2}\right.\non\\
&&\left.+2C(n) \zeta^{*}(n)\zeta(n)+2D(n)\zeta^{*}(n)\zeta(n-1)\right\},
\label{app-fluc-action}
\eeqa
where
\sbeqa
\label{ABCD}
\beqa
&&A(n) = \epsilon\Omega\cos 2\phi,\\
&&B(n) = \epsilon\Omega\cos 2\phi,\\
&&C(n) = -i, \\
&&D(n) = i -\epsilon\Omega\cos 2\phi.
\eeqa
\seeqa 
This can be expressed in the matrix form:
\beqa
\frac{i}{\hbar}{\cal S}^{{\rm ss}}_{2}(N-1) = -i\frac{N_{{\rm DW}}S}{2}~ ^t\mbox{\boldmath$\zeta$}\mbox{\boldmath$M$}(N-1)\mbox{\boldmath$\zeta$},
\eeqa
where
\sbeqa
\beqa
 ^t \mbox{\boldmath$\zeta$}&:=& \left( \zeta^{*}(N-1),\zeta(N-1),\zeta^{*}(N-2),...,\zeta^{*}(1),\zeta(1) \right) \\
&&\non\\
\mbox{\boldmath$M$}(N-1) &:=&
\left(
	\begin{array}{cccccc}
A(N-1) & C(N-1) & 0      &        &       &      \\
C(N-1) & B(N-1) & D(N-1) &\ddots  &       &      \\
 0     & D(N-1) & A(N-2) &\ddots  &\ddots &      \\
       &        & \ddots &\ddots  & D(2)  &      \\
       &        &        & D(2)   & A(1)  &0     \\ 
       &        &        &   0    & C(1)  & B(1) 
\end{array}\right).
\eeqa
\seeqa
As to the integration measure, 
which has not been explicitly derived, 
we may make the following two assumptions:

(i) Its structure is the same as that of the (boson-)coherent-state path integral:
\beqa
\frac{1}{{\cal M}}\frac{d\zeta(n)d\zeta^{*}(n)}{2\pi i},
\eeqa
where ${\cal M}$ is a constant.

(ii) The value of ${\cal M}$ can be 
inferred from the overlap between the initial and the final states.

Assumption (ii) is reasonable since 
$\exp[i{\cal S}^{{\rm ss}}(T=0)/\hbar]$ should coincide with $\<z_{{\rm F}}|z_{{\rm I}}\>$. 
On the basis of these assumptions, 
the fluctuation integral can be cast into the form
\beqa
K_{2}(T) &=& 
\lim_{N\to\infty}\int\prod_{n=1}^{N-1}\frac{1}{{\cal M}}
\frac{d\zeta(n)d\zeta^{*}(n)}{2\pi i}\exp\left[-i\frac{N_{{\rm DW}}S}{2}
~ ^t\mbox{\boldmath$\zeta$}{\bf \mbox{\boldmath$M$}}(N-1)\mbox{\boldmath$\zeta$} \right]\non\\
&=& \lim_{N\to\infty}\left[
({\cal M}N_{{\rm DW}}S)^{2(N-1)}(-1)^{N-1}
\det(i\mbox{\boldmath$M$}(N-1))\right]^{-1/2}\non\\
&=& \lim_{N\to\infty}\left[({\cal M}N_{{\rm DW}}S)^{2(N-1)}\det\mbox{\boldmath$M$}(N-1)\right]^{-1/2}.
\eeqa
The above determinant can be evaluates as follows.\cite{Solari} 
Let 
\beqa
M(n) &:=& \det\mbox{\boldmath$M$}(n)=
\left|
	\begin{array}{cccc}
A(n) & -i   & 0     & \\
-i   &  B(n)& D(n)  & \\
0    & D(n) & A(n-1)& \\
     &      &       &\ddots
 \end{array}\right|.
\eeqa
This can be expanded in terms of the cofactor as
\beqa
M(n)= A(n)M'(n)+M(n-1),
\label{rec1}
\eeqa
where 
\beqa
M'(n) &:=& 
\left|
	\begin{array}{cccc}
 B(n)&  D(n)   &0      & \\
 D(n)& A(n-1)  &-i     & \\
0    &-i       & B(n-1)& \\
     &         &       &\ddots
 \end{array}\right|.
\eeqa
$M'(n)$ can in turn be expanded as 
\beqa
M'(n) = B(n)M(n-1)-D^{2}(n)M'(n-1).
\label{rec2}
\eeqa
The recursion relations (\ref{rec1}) and (\ref{rec2}) 
should be solved 
with the initial condition 
\beqa
M(0)=1,\qquad M'(0)=0.
\eeqa
In the limit of $\epsilon \to 0$, 
they reduce to a set of coupled first-order differential equations:
\sbeqa
\beqa
&&\frac{dM(t)}{dt} = \Omega_{\phi}M'(t),\\
&&\frac{dM'(t)}{dt} = \Omega_{\phi}\left\{M(t) + 2iM'(t)\right\},\\
&& M(0)=1,\qquad M'(0)=0,\non
\eeqa
\seeqa
where $\Omega_{\phi}$ is given by (\ref{K2T}). 
This gives 
\beqa
M(t) = (1-i\Omega_{\phi}t)e^{i\Omega_{\phi}t},\qquad 
M'(t) = \Omega_{\phi}t e^{i\Omega_{\phi}t}.
\eeqa
Hence, 
\beqa
K_{2}(T) &=& \lim_{N\to\infty}\left[({\cal M}N_{{\rm DW}}S)^{2(N-1)}M(N-1)\right]^{-1/2} \non\\
&=& \frac{e^{-i\Omega_{\phi}T/2}}{\sqrt{1-i\Omega_{\phi}T}}\lim_{N\to\infty}\frac{1}{({\cal M}N_{{\rm DW}}S)^{N-1}}.
\eeqa
Since $K_{2}(0)$ should be unity, we conclude from 
assumption (ii) that 

\beqa
{\cal M} = \frac{1}{N_{{\rm DW}}S}.
\eeqa




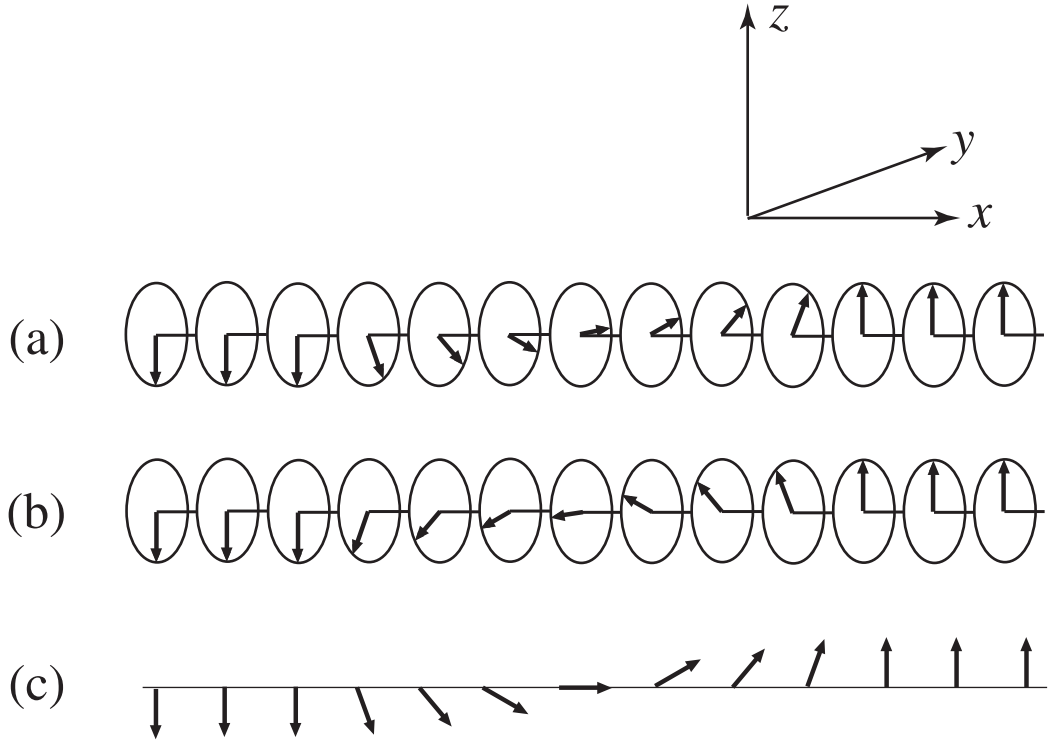
\begin{figure}[htbp]
\caption[Fig. 1]{Domain walls with three chiralities (quoted from Ref. 10); (a) right-handed wall $(\phi_{0}=\pi/2)$, (b)left-handed wall $(\phi_{0}=-\pi/2)$, and (c) wall with no chirality $(\phi_{0}=0)$. Circles in (a) and (b) drawn to guide the eye lie in the $yz$ plane, while the spins lie in the $zx$ plane in (c). The quasi-one-dimensional direction of the crystal is here aligned with the spin hard axis for ease of visualization. A different alignment, which may be the case for a real magnet, does not affect the content of the text; for instance, one could rotate all the spins by $\pi/2$ around the $y$ axis if the dominant anisotropy originates from the demagnetizing field. }
\label{Fig.1}
\end{figure}

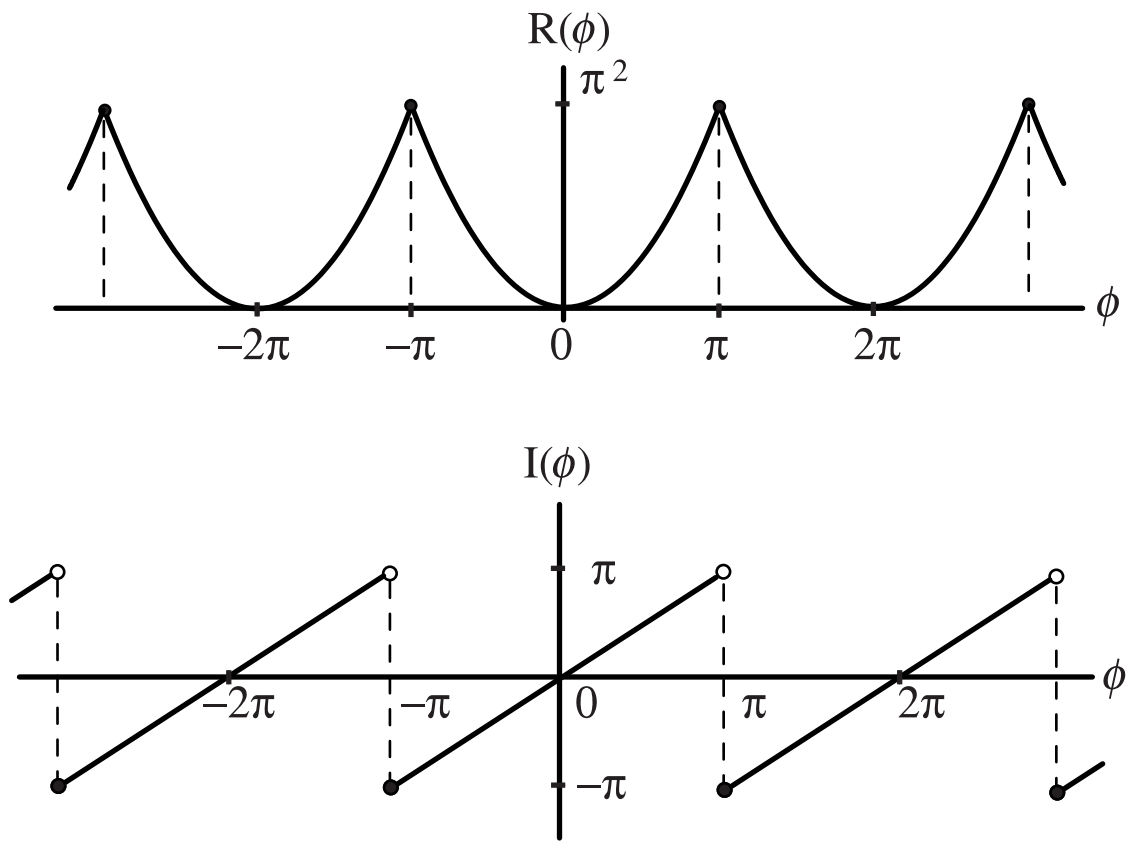
\begin{figure}[htbp]
\caption[Fig. 2]{Functions $R(\phi)$ and $I(\phi)$.~~~~~~~~~~~~~~~~~~~~~~~~~~~~~~~~~~~~~~~~~~~~~~~~~~~~~~~~~~~~~~~~~~~~~~~~~~}
\label{Fig.2}
\end{figure}

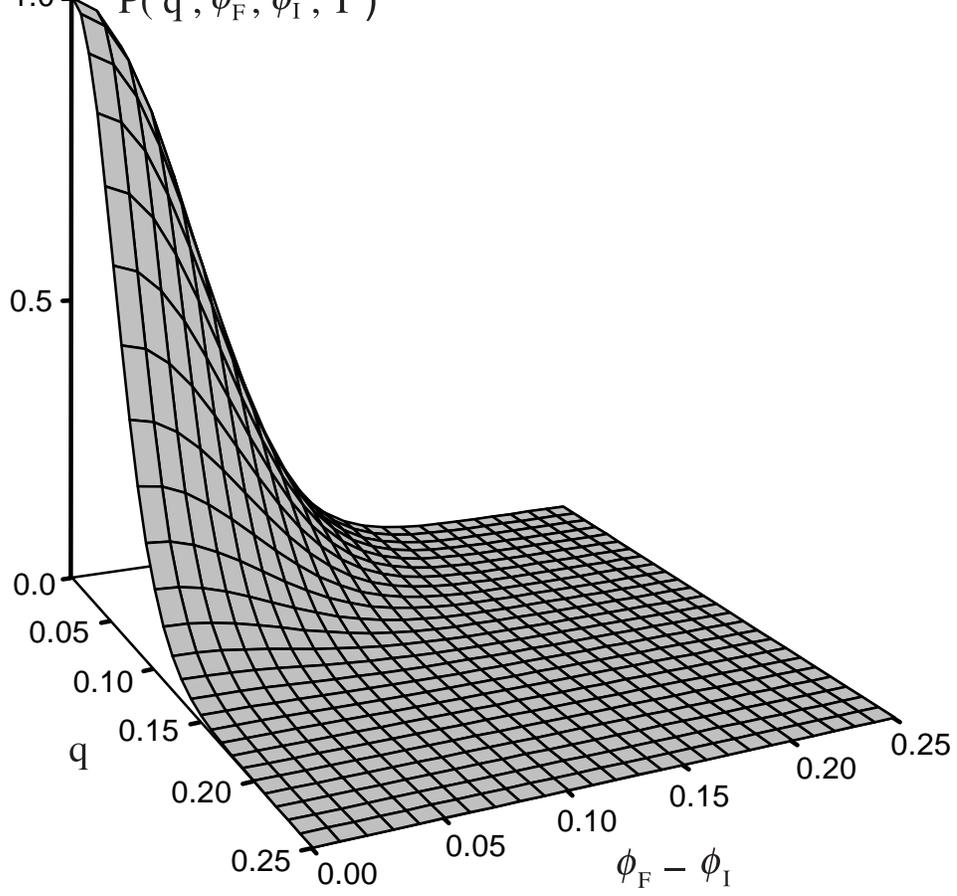
\begin{figure}[htbp]
\caption[Fig. 3]{The transition probability 
in the absence of transverse anisotropy as a function of 
$q\equiv q_{{\rm F}}-q_{{\rm I}}$ 
and $\phi_{{\rm F}}-\phi_{{\rm I}}$ with $N_{{\rm DW}}S=100$. 
It is independent of $T$.}
\label{Fig.3}
\end{figure}

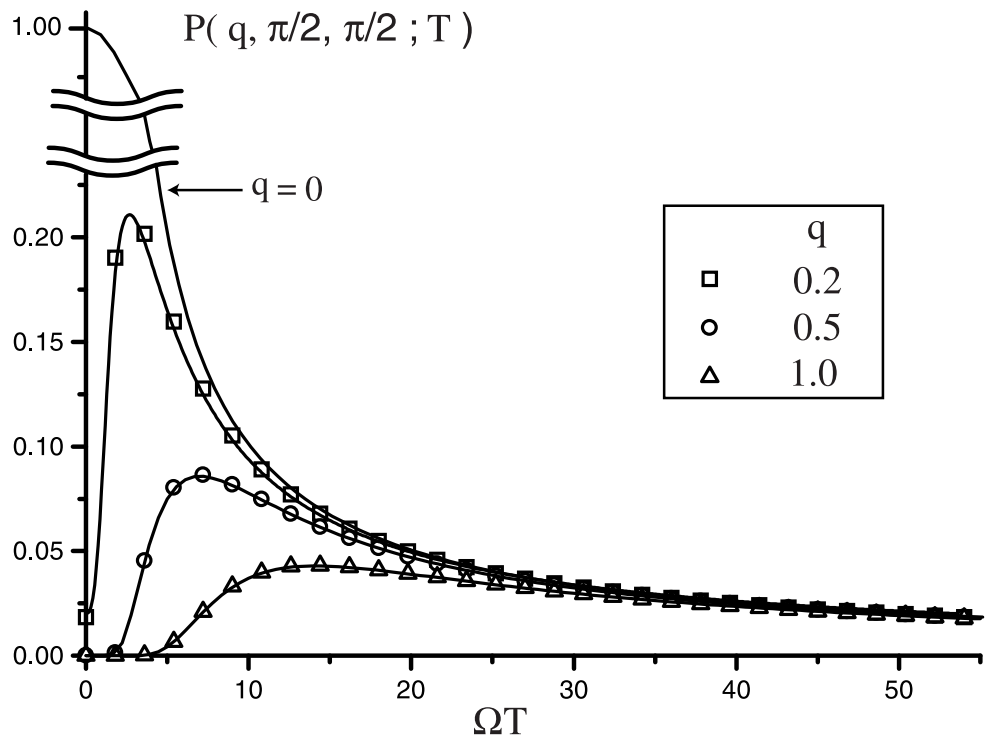
\begin{figure}[htbp]
\caption[Fig. 4]{Time-dependence of the transition probabilities 
for $\phi_{{\rm F}}=\phi_{{\rm I}}=\pi/2$. 
Solid lines represent analytical results in the linear approximation 
(the curve for $q=0$ is exact). 
Squares, open circles, and triangles represent 
numerical results.}
\label{Fig.4}
\end{figure}

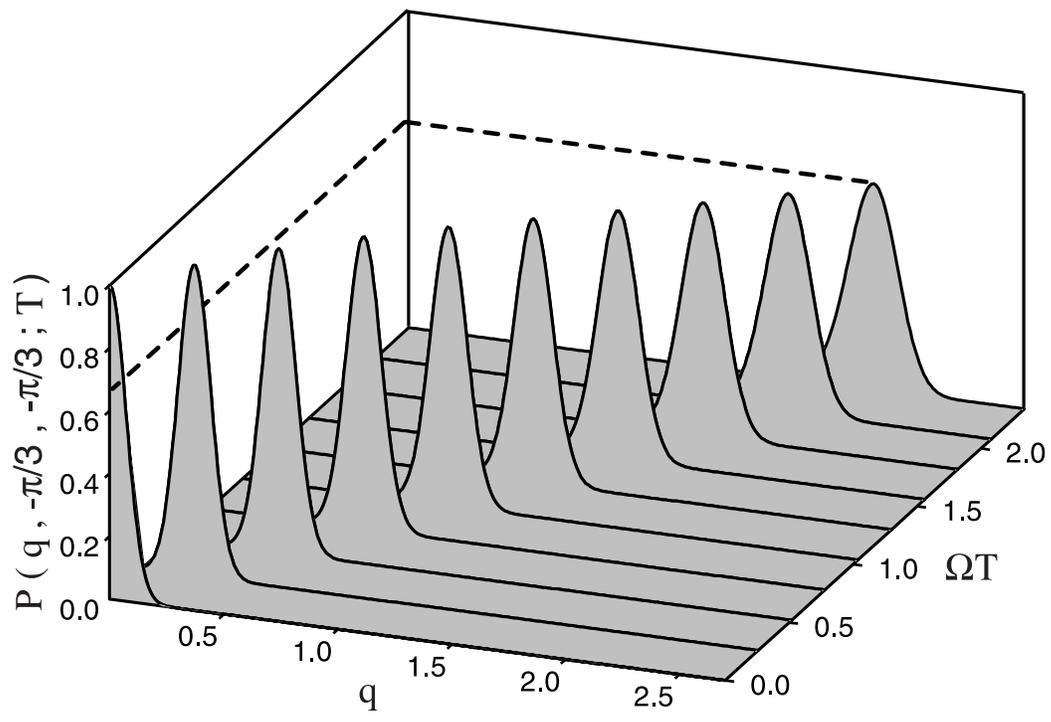
\begin{figure}[htbp]
\caption[Fig. 5]{The transition probability 
as a function of $q$ and $T$ in the case of 
$\phi_{{\rm F}}=\phi_{{\rm I}}=-\pi/3$. 
Numerical results are indistinguishable from the analytical ones.}
\label{Fig.5}
\end{figure}

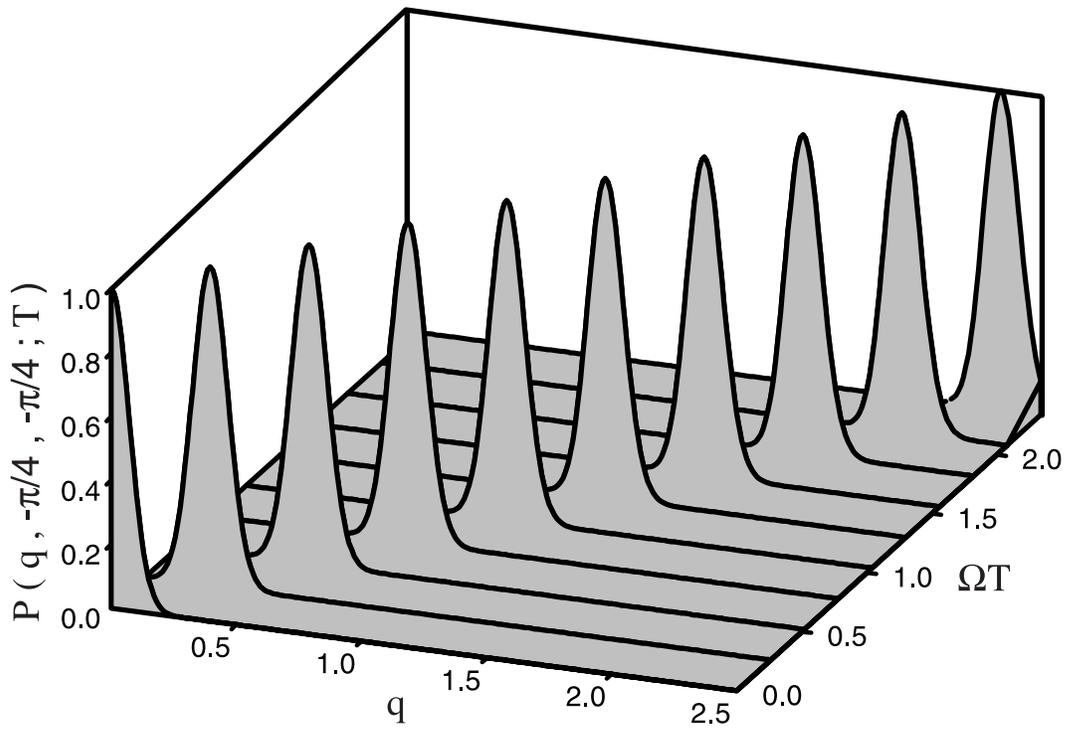
\begin{figure}[htbp]
\caption[Fig. 6]{The transition probability 
as a function of $q$ and $T$ in the case of 
$\phi_{{\rm F}}=\phi_{{\rm I}}=-\pi/4$. 
Numerical results are indistinguishable from the analytical ones.}
\label{Fig.6}
\end{figure}

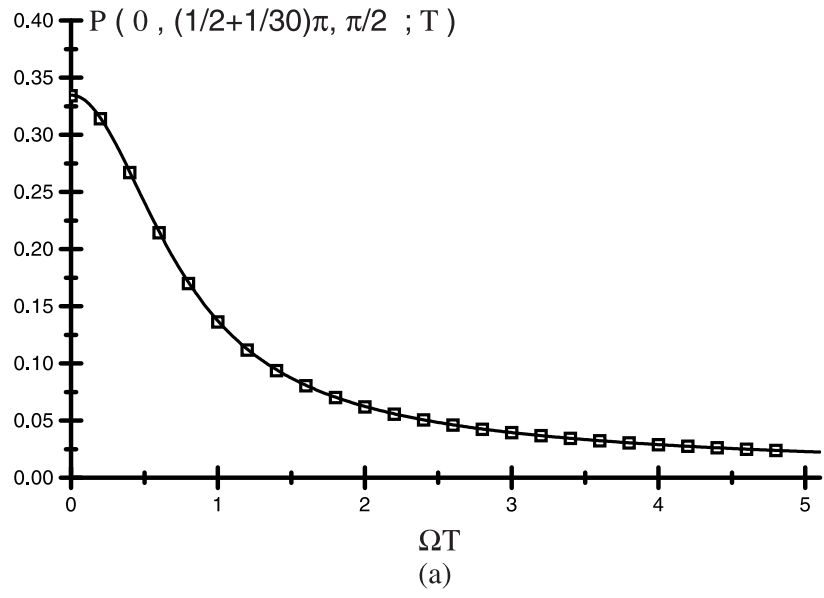
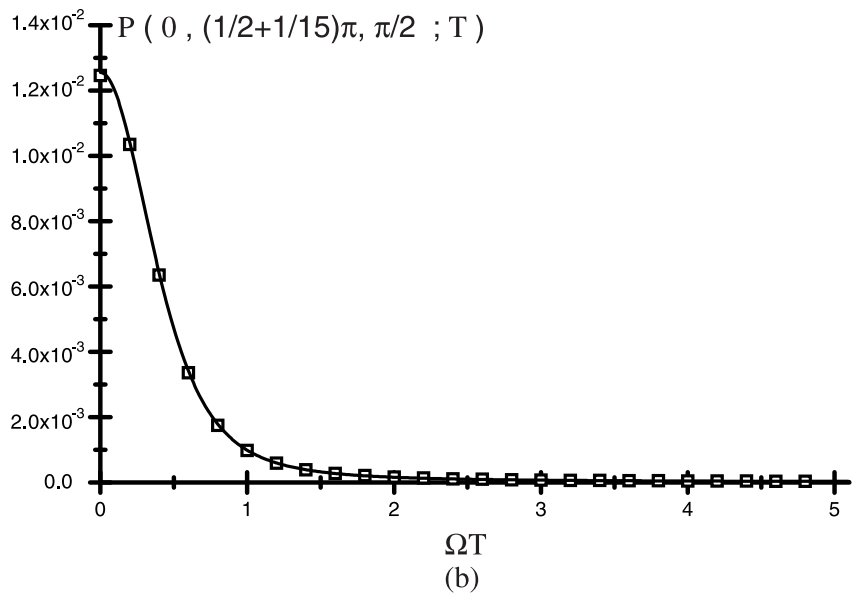
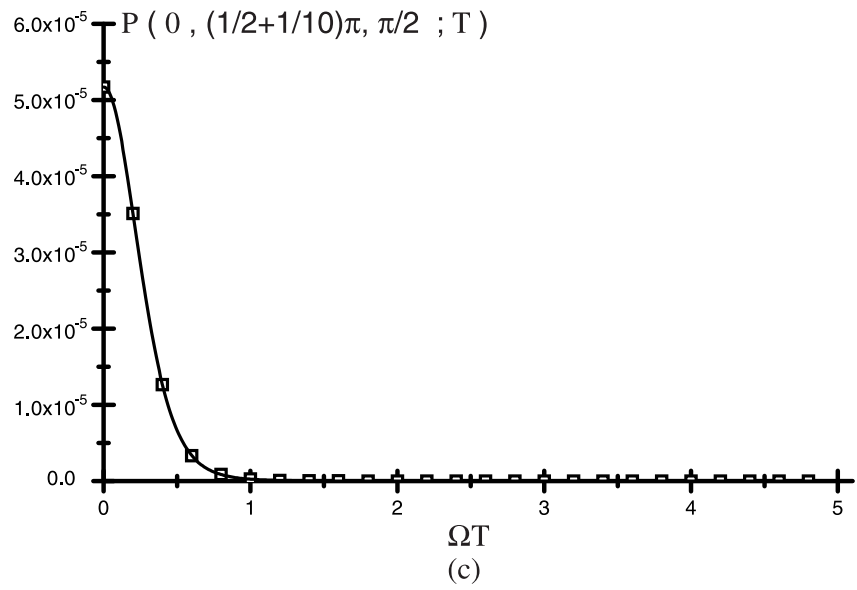
\begin{figure}[htbp]
\caption[Fig. 7]{Time-dependence of the transition probabilities for 
$q_{{\rm F}}=q_{{\rm I}}$ and $\phi_{{\rm I}}=\pi/2$. 
Solid lines represent analytical results in the linear approximation. 
Squares represent numerical results.}
\label{Fig.7}
\end{figure}


\begin{references}
\bibitem[\dag]{emailadd}email: shibata@cmpt01.phys.tohoku.ac.jp
\bibitem[\dag\dag]{emailadd}email: takagi@cmpt01.phys.tohoku.ac.jp

\bibitem{S-C-B1}
P. C. E. Stamp, E. M. Chudnovsky, and B. Barbra, Int. J. Mod. Phys. B {\bf 6}, 1355 (1992).

\bibitem{QTM94}
{\it Quantum Tunneling of Magnetization}, Proceedings of the NATO workshop, Chichilianne, France, 1994, edited by L. Gunther and B. Barbara (Kluwer Academic, Norwell, MA, 1995).

\bibitem{C-T}
E. M. Chudnovsky and J. Tejada, {\it Macroscopic Quantum Tunneling of the Magnetic Moment} (Cambridge University Press, United Kingdom, 1998).


\bibitem{Egami}
T. Egami, Phys. Status Solidi B {\bf 57}, 211 (1973); Phys. Status Solidi A {\bf 19}, 747 (1973); {\bf 20}, 157 (1973).

\bibitem{Stamp}
P. C. E. Stamp, Phys. Rev. Lett. {\bf 66}, 2802 (1991).

\bibitem{Chudnovsky}
E. M. Chudnovsky, O. Iglesias, and P. C. E. Stamp, Phys. Rev. B {\bf 46}, 5392 (1992).

\bibitem{Tatara-Fukuyama}
G. Tatara and H. Fukuyama, Phys. Rev. Lett. {\bf 72}, 772 (1994); J. Phys. Soc. Jpn. {\bf 63}, 2538 (1994).

\bibitem{Braun-Loss}
H. B. Braun and D. Loss, Phys. Rev. B {\bf 53}, 3237 (1996).

\bibitem{Braun-Loss2}
H. B. Braun and D. Loss, J. Appl. Phys. {\bf 79}, 6107 (1996).

\bibitem{Takagi-Tatara}
S. Takagi and G. Tatara, Phys. Rev. B {\bf 54}, 9920 (1996).

\bibitem{Ivanov}
B. A. Ivanov, A. K. Kolezhuk, and V. E. Kireev, Phys. Rev. B {\bf 58}, 11514 (1998).


\bibitem{Caldeira-Leggett}
A. O. Caldeira and A. J. Leggett, Ann. Phys. (N.Y.) {\bf 149}, 347 (1983).



\bibitem{Leggett}
A. J. Leggett, Prog. Theor. Phys. Suppl. {\bf 69}, 80 (1980).

\bibitem{Klauder}
J. R. Klauder, Phys. Rev. D {\bf 19}, 2349 (1979).

\bibitem{Solari}
H. G. Solari, J. Math. Phys. {\bf 28}, 1097 (1987).

\bibitem{Kochetov}
E. A. Kochetov, J. Math. Phys. {\bf 36}, 4667 (1995).

\bibitem{Shibata-Takagi}
J. Shibata and S. Takagi, Int. J. Mod. Phys. B {\bf 13}, 107 (1999).

\bibitem{Radcliffe}
J. M. Radcliffe, J. Phys. A {\bf 4}, 313 (1971).

\bibitem{Rajaraman}
R. Rajaraman, {\it Solitons and Instantons} (North-Holland, Amsterdam, 1982).

\bibitem{Abramowitz-Stegun}
{\it Handbook of Mathematical Functions with Formulas, Graphs, and Mathematical Tables}, edited by M. Abramowitz and I. A. Stegun (Dover Publications, Inc., New York, 1965).

\end{references}
\end{document}